\documentclass{article}
\usepackage{amsmath,amsfonts,setspace,cancel,url,hyperref,verbatim,color,todonotes}
\usepackage{caption} 
\usepackage{cite}
\usepackage{stmaryrd} 
\usepackage[a4paper]{geometry}
\usepackage{breqn}
\usepackage{tikz}
\usepackage{tikz-cd}
\newcommand\Tstrut{\rule{0pt}{3ex}}         
\newcommand\Bstrut{\rule[-1.3ex]{0pt}{0pt}}   

\newcommand{\SL}[1]{\mathrm{SL}( #1 )}
\newcommand{\SO}[1]{\mathrm{SO}( #1 )}
\newcommand{\GL}[1]{\mathrm{GL}( #1 )}
\newcommand{\ODD}{\mathrm{O}(D,D)}
\newcommand{\SLf}{\SL{5}}

\newcommand{\G}{\GL{5}^+}
\newcommand{\Gp}{\SLf\times\mathbb{R}^+}

\newcommand{\Edd}{E_{d(d)}}
\newcommand{\DWZW}{DFT_{WZW}}

\newcommand{\gL}{\mathcal{L}} 
\newcommand{\gE}{\mathbb{E}} 
\newcommand{\e}{E} 
\newcommand{\tE}{\tilde{\gE}} 
\newcommand{\gM}{{\cal M}} 
\newcommand{\tnabla}{\tilde{\nabla}} 
\newcommand{\tGamma}{\tilde{\Gamma}} 
\newcommand{\Gammao}{\mathring{\Gamma}} 
\newcommand{\tS}{\tilde{S}} 
\newcommand{\tZ}{\tilde{Z}} 
\newcommand{\tT}{T} 

\newcommand{\IM}{d^{10}Y d^7x\, \gE e\,}
\newcommand{\Conn}{\omega}

\newcommand{\hpartial}{\hat{\partial}}

\newcommand{\TAw}{{\cal A}(1/5)}
\newcommand{\TBw}{{\cal B}(2/5)}
\newcommand{\TCw}{{\cal C}(3/5)}
\newcommand{\TDw}{{\cal D}(4/5)}
\newcommand{\TSw}{{\cal S}(1)}
\newcommand{\TAwl}{{\cal A}(1/5,0)}
\newcommand{\TBwl}{{\cal B}(2/5,1/2)}
\newcommand{\TCwl}{{\cal C}(3/5,1/2)}
\newcommand{\TDwl}{{\cal D}(4/5,1)}
\newcommand{\TSwl}{{\cal S}(1,1)}
\newcommand{\TA}{{\cal A}}
\newcommand{\TB}{{\cal B}}
\newcommand{\TC}{{\cal C}}
\newcommand{\TD}{{\cal D}}

\newcommand{\Aa}{{\TA}}
\newcommand{\Ab}{{\TB}}
\newcommand{\Ac}{{\TC}}
\newcommand{\Ad}{{\TD}}
\newcommand{\Fa}{\mathcal{F}}
\newcommand{\Fb}{\mathcal{H}}
\newcommand{\Fc}{\mathcal{J}}
\newcommand{\Fd}{\mathcal{K}}

\newcommand{\D}{\mathfrak{D}}

\newcommand{\bmu}{\bar{\mu}}

\newcommand{\bba}{\bar{a}}
\newcommand{\bbb}{\bar{b}}
\newcommand{\bbc}{\bar{c}}
\newcommand{\bbd}{\bar{d}}
\newcommand{\bee}{\bar{e}}

\newcommand{\be}{\begin{equation}}
\newcommand{\ee}{\end{equation}}

\numberwithin{equation}{section}
\begin{document}

\begin{titlepage}
\vfill

\begin{flushright}
MPP-2016-94 \\
LMU-ASC 19/16
\end{flushright}

\vfill

\begin{center}
   \baselineskip=16pt
   	{\Large \bf A geometric formulation of exceptional field theory}
   	\vskip 2cm
	{\bf Pascal du Bosque$^{a,b}$\footnote{\tt dubosque@mpp.mpg.de}, Falk Hassler$^{c,d,e}$\footnote{\tt fhassler@unc.edu}, Dieter L\"ust$^{a,b}$\footnote{\tt dieter.luest@lmu.de}, Emanuel Malek$^a$\footnote{\tt e.malek@lmu.de}}
	\vskip .6cm
    {\small \it $^a$ Arnold Sommerfeld Center for Theoretical Physics, Department f\"ur Physik, \\ Ludwig-Maximilians-Universit\"at M\"unchen, Theresienstra{\ss}e 37, 80333 M\"unchen, Germany \\ \ \\
    $^b$ Max-Planck-Institut f\"ur Physik, Werner-Heisenberg-Institut, \\ F\"ohringer Ring 6, 80805 M\"unchen, Germany \\ \ \\
    $^c$ Department of Physics and Astronomy, University of North Carolina, \\
    Phillips Hall, CB \#3255, 120 E. Cameron Ave., Chapel Hill, NC 27599-3255, USA \\ \ \\
    $^d$ City University of New York, The Graduate Center, \\
    365 Fifth Avenue, New York, NY 10016, USA \\ \ \\
    $^e$ Department of Physics, Columbia University, \\
    Pupin Hall, 550 West 120th St., New York, NY 10027, USA} \\
	\vskip 2cm
\end{center}

\begin{abstract}
We formulate the full bosonic $\SLf$ exceptional field theory in a coordinate-invariant manner. Thereby we interpret the 10-dimensional extended space as a manifold with $\Gp$-structure. We show that the algebra of generalised diffeomorphisms closes subject to a set of closure constraints which are reminiscent of the quadratic and linear constraints of maximal seven-dimensional gauged supergravities, as well as the section condition. We construct an action for the full bosonic $\SLf$ exceptional field theory, even when the $\Gp$-structure is not locally flat.

\end{abstract}

\vfill

\setcounter{footnote}{0}
\end{titlepage}

\tableofcontents

\newpage

\section{Introduction}
Double field theory (DFT) \cite{Hull:2009mi,Hull:2009zb,Hohm:2010jy,Hohm:2010pp} has grown out of a desire to better understand T-duality by using a formalism in which it is made manifest \cite{Siegel:1993th,Siegel:1993xq,Tseytlin:1990nb,Duff:1989tf}. It does so at the level of the target-space action by doubling the number of coordinates and introducing the generalised Lie derivative that generates the local symmetries of the theory. For the toroidal case the extra coordinates can be understood from string field theory \cite{Hull:2009mi} as being dual to momenta and winding modes. However, the physical fields of the theory do not depend on all coordinates. They are subject to a constraint, known as the ``section condition'', which restricts their dependence to a maximal isotropic subspace of the coordinates \cite{Hull:2009mi,Siegel:1993th}. Using this constraint, the action of double field theory reduces to that of type II supergravity, the generalised Lie derivative generates diffeomorphisms and $B$-field gauge transformations and the resulting formulation looks reminiscent of generalised geometry \cite{Gualtieri:2003dx,Hitchin:2004ut,Coimbra:2011nw}.

The generalisation of double field theory to U-dualities, known as exceptional field theory (EFT) \cite{Berman:2010is,Berman:2011cg,Berman:2011pe,Berman:2011jh,Berman:2012vc,Hohm:2013vpa,Berman:2013eva,Hohm:2013uia,Hohm:2014fxa,Godazgar:2014sla,Godazgar:2014nqa,Musaev:2014lna,Cederwall:2015ica,Berman:2015rcc,Musaev:2015ces} \footnote{See also \cite{Hillmann:2009ci} for independent but related work.}, uses an ``extended coordinate space'' which grows quickly with the rank of the U-duality group, $\Edd$. The closure of its algebra of generalised diffeomorphisms also requires a section condition \cite{Berman:2011cg,Berman:2012vc,Coimbra:2012af} which in turn restricts the coordinate dependence of physical fields. This condition has two inequivalent solutions \cite{Hohm:2013vpa,Blair:2013gqa} and depending on which one is used, the action reduces to that of 11-dimensional supergravity or type IIB supergravity. With the section condition, the resulting formulation now resembles \emph{exceptional} generalised geometry \cite{Coimbra:2012af}.

One of the interesting problems in the field is to determine what the geometry underlying the extended space is. This is pertinent if one wants to better understand non-geometric flux compactifications for which de Sitter no-go theorems may not apply \cite{deCarlos:2009qm,deCarlos:2009fq,Blaback:2013ht,Damian:2013dq,Damian:2013dwa,Hassler:2014mla}. For example, one may wish to generalise the powerful results of \cite{Ashmore:2015joa,Ashmore:2016qvs} to include non-geometry.

Interesting progress has been made on this front by studying the exponentiation of the local symmetries of the theories \cite{Hohm:2012gk,Park:2013mpa,Hohm:2013bwa, Berman:2014jba,Hull:2014mxa,Papadopoulos:2014mxa,Naseer:2015tia,Rey:2015mba,Chaemjumrus:2015vap}. One hope is that these transformations can be used to patch the extended spaces. However, these proposals are either defined on section, as in \cite{Hull:2014mxa,Rey:2015mba,Chaemjumrus:2015vap} or already make use of a flat structure  \cite{Hohm:2012gk,Park:2013mpa,Hohm:2013bwa,Berman:2014jba,Naseer:2015tia}.

Here we try and understand the extended space as a manifold with reduced structure group. We make this explicit for the case of the $\SLf$ EFT relevant to seven-dimensional compactifications and show that one can define a ``curved'' exceptional field theory for any 10-dimensional manifold with $\G$-structure\footnote{$\G \simeq \Gp$ in our notation}, not just for locally flat $\G$-structures, as would be required for the usual EFT formulation. Our formulation resembles that of ``DFT on group manifolds'' \cite{Blumenhagen:2014gva,Blumenhagen:2015zma,Bosque:2015jda}, which we will henceforth refer to as $\DWZW$. However, the crucial difference is that we \emph{do not} require the vielbein to be the Maurer-Cartan form of the gauge group described by the background. The work here also shares many ideas with \cite{Cederwall:2014kxa,Cederwall:2014opa}, whilst extending them to EFT.

We will begin with a review of the essential features of $\SLf$ exceptional field theory in section \ref{s:Review} before discussing the geometry of 10-dimensional manifolds with $\G$-structure in section \ref{s:Geometry}. We define the relevant $\G$-structure, use it to construct the generalised Lie derivative and show that requiring the algebra to close leads to a set of constraints, including the section condition. We also discuss the case when the $\G$-structure is locally flat and show how this case reduces to the usual EFT. In section \ref{s:TH} we develop the formalism in order to describe the tensor hierarchy of the EFT, closely following \cite{Hohm:2015xna,Wang:2015hca}. Finally, in section \ref{s:Action} we give the full action, including the ``external'' seven-dimensional fields, for any $\G$-structure, not just locally flat ones. The resulting theory is manifestly coordinate invariant, as well as invariant under generalised diffeomorphisms and external diffeomorphisms. We discuss background-dependence and comment on further work in section \ref{s:Discussion}.

\section{A brief review of $\SLf$ exceptional field theory} \label{s:Review}
Let us briefly review exceptional field theory, focusing on the $\SLf$ EFT \cite{Berman:2010is,Berman:2011cg,Blair:2013gqa,Musaev:2015ces} which, for example, can be used to describe seven-dimensional maximal gauged supergravities. The theory has 10 ``extended coordinates'' $Y^{[ab]}$ and seven ``external coordinates'' $x^\mu$, with $a, b = 1, \ldots, 5$ and $\mu = 1, \ldots, 7$. The bosonic degrees of freedom of the internal sector are described by a generalised metric
\begin{equation}
 \gM_{ab} \in \SLf/\mathrm{SO}(5) \,,
\end{equation}
which can be parameterised by a four-dimensional metric and three-form corresponding to the internal sector of 11-dimensional supergravity, or by a three-dimensional metric, a doublet of two-forms and 3 scalars, the internal sector of IIB \cite{Blair:2013gqa}.

Just as the bosonic degrees of freedom can be unified in the generalised metric $\gM_{ab}$, its symmetries, corresponding to diffeomorphisms and $p$-form gauge transformations, are combined into the generalised Lie derivative
\begin{equation}
 \gL_{\Lambda} V^a = \frac{1}{2} \Lambda^{bc} \partial_{bc} V^a + \frac15 V^a \partial_{bc} \Lambda^{bc} - V^a \partial_{bc} \Lambda^{ac} \,. \label{eq:RevGenLie}
\end{equation}
Here the parameter of generalised diffeomorphisms, $\Lambda^{ab}$ , has weight $\frac15$ under the generalised Lie derivative, so that under a generalised Lie derivative it transforms as
\begin{equation}
 \gL_{\Lambda_1} \Lambda_2^{ab} = \frac12 {\Lambda_1}^{cd} \partial_{cd} \Lambda_2^{ab} + \left( \frac25 + \frac{1}{10} \right) \Lambda_2^{ab} \partial_{cd} \Lambda_1^{cd} - \Lambda_2^{cb} \partial_{cd} \Lambda_1^{ad} - \Lambda_2^{ac} \partial_{cd} \Lambda_1^{bd} \,.
\end{equation}

In order for these transformations to close into an algebra
\begin{equation}
 \left[ \gL_{\Lambda_1}, \gL_{\Lambda_2} \right] V^a = \gL_{[\Lambda_1, \Lambda_2]_E} V^a \,, \label{eq:RevClosure}
\end{equation}
where the E-bracket is the antisymmetrisation of the generalised Lie derivative, i.e.
\begin{equation}
 \left[ \Lambda_1, \Lambda_2 \right]_E = \frac12 \left( \gL_{\Lambda_1} \Lambda_2 - \gL_{\Lambda_2} \Lambda_1 \right) \,,
\end{equation}
one imposes the so-called ``section condition''
\begin{equation}
 \partial_{[ab} f\, \partial_{cd]} g = 0 \,, \qquad \partial_{[ab} \partial_{cd]} f = 0\,, \label{eq:RevSC}
\end{equation}
when acting on any pair of fields $f$ and $g$. There are two inequivalent solutions (i.e. not related by $\SLf$ transformations) to the section condition \cite{Blair:2013gqa,Hohm:2013vpa}, given by
\begin{equation}
 \begin{split}
 (i)\phantom{ii} \quad \partial_{ij} &= 0 \,, \textrm{ where } i = 1, \ldots, 4 \,, \\
 (ii) \quad \partial_{A\alpha} &= \partial_{\alpha\beta} = 0 \,, \textrm{ where } \alpha, \beta = 4, 5 \,, \; A, B = 1, \ldots, 3\,.
 \end{split}
\end{equation}
The first, where fields only depend on the four coordinates $Y^{i5}$ corresponds to 11-dimensional supergravity while the second, with dependence on only three coordinates $Y^{AB}$, corresponds to type IIB supergravity. One way this manifests itself is that the generalised Lie derivative of the generalised metric generates exactly the diffeomorphisms and $p$-form gauge transformations of the bosonic fields of 11-dimensional / IIB supergravity.

Furthermore, one can define a unique action which is invariant under generalised Lie derivatives. Upon using the appropriate solution of the section condition this reduces to the four-dimensional / three-dimensional internal sector of 11-dimensional / IIB supergravity \cite{Berman:2010is,Blair:2013gqa}. This can be extended by introducing fields related to the tensor hierarchy of gauged supergravities \cite{deWit:2005hv,deWit:2008ta}, so that the resulting action reduces to the bosonic part of the full 11-dimensional or IIB supergravity \cite{Hohm:2013uia,Hohm:2013vpa,Musaev:2015ces}. We will describe this construction in more detail in sections \ref{s:TH}, \ref{s:Action} and appendix \ref{A:ExtDiffeos}, albeit in our geometric formulation.

\section{Structure group, generalised Lie derivative and $\GL{10}$ connection} \label{s:Geometry}

\subsection{10-manifolds with $\G$-structure}
\subsubsection{$\G$-structure}
In this paper we define a ``curved'' version of the $\SL{5}$ EFT on a 10-dimensional manifold $M$ which admits a $\G$-structure. In order to do this, consider first the usual frame bundle $\pi_{10}: F_{10}M \longrightarrow M$ whose fibre consists of all ordered bases of the tangent bundle and can thus be identified with $\GL{10}$. We will label the bases as $\gE_{ab}$. The structure group of $M$ can be reduced to $\G$ if $F_{10}M/\G$ admits a global section and each such global section defines a $\G$-structure on $M$. In other words, a $\G$-structure is defined by an equivalence class of frame fields $\gE_{ab}$,
\begin{equation}
 \gE_{ab} \sim \gE'_{ab} \iff \gE_{ab} = u_{a}{}^c u_b{}^d \gE'_{cd} \,, \label{eq:EquRel}
\end{equation}
where $u \in \G$. Here $a$, $b = 1, \ldots, 5$ and the pair of indices $ab$ is antisymmetrised, thus denoting the 10-dimensional representation of $\G$. In local coordinates we will write the frame fields as 
\begin{equation}
 \gE_{ab} = \gE_{ab}{}^M \partial_M \,,
\end{equation}
with $M = 1, \ldots, 10$ denoting ``curved'' 10-dimensional indices. A global section of $F_{10}M / \G$ then implies that the transition functions of the frame bundle can be chosen to be $\G$-valued. For each $\G$-structure, we can define a principal $\G$-bundle $\pi_5: F_5M \longrightarrow M$, whose fibres consist of the equivalence class of frame fields defining the $\G$-structure and can thus be identified with $\G$.

Note that the $\G$-structure can also be defined using an invariant tensor. The product $\mathbf{10} \otimes \mathbf{10} \otimes \overline{\mathbf{10}} \otimes \overline{\mathbf{10}}$ of $\GL{10}$ contains a singlet in the decomposition under $\G$, corresponding to a $\G$-invariant tensor, $Y^{ab,cd}_{ef,gh} = \epsilon^{abcdi} \epsilon_{efghi}$, the ``$Y$-tensor'' in the nomenclature of \cite{Berman:2012vc}. However, here we will find it more useful to use the equivalence class of frame-fields \eqref{eq:EquRel} instead of the $Y$-tensor when discussing the $\G$-structure.

Unlike in the usual EFT formulation, we do not assume that our $\G$-structure is locally flat. The obstruction to local flatness of this structure introduces a manifest background-independence into our theory. We will return to this point briefly in the discussion \ref{s:Discussion}. Finally, let us emphasise that in general the representative of a $\G$-structure \emph{cannot} be written in the form
\begin{equation}
 \gE_{ab}{}^{M} \longarrownot\longrightarrow \gE_{ab}{}^{ij} = \e_{[a}{}^i \e_{b]}{}^j \,, \label{eq:CanStructure}
\end{equation}
for some $\e_a{}^i$, where $i, \, j = 1, \ldots, 5$. When this can be done, the $\G$-structure is called locally flat.

\subsubsection{Fundamental vector fields}

Equipped with $F_5M$ we can construct a 5-dimensional vector bundle, the associated bundle $E_5$ on which $\G$ acts in the fundamental representation. The sections of this bundle are ``fundamental vector fields'' with basis $\e_a$, so that we can write
\begin{equation}
 V = V^a \e_a \,, \textrm{ for } V \in E_5 \,.
\end{equation}

Because the vector bundle $E_5$ has structure group $\G$, we can define a ``5-dimensional volume-form'', $\eta$, as a global section of $\Lambda^5 E_5^*$. In the $\e_a$ basis we denote this by
\begin{equation}
 \eta_{abcde} = |\gE|^{1/2} \epsilon_{abcde} \,,
\end{equation}
where $|\gE|$ is the determinant of $\gE_{ab}{}^M$ and $\epsilon_{abcde}$ is the alternating symbol which equals the sign of the permutation $\left(abcde\right)$. We will often find it useful to use the tensor density $\epsilon_{abcde}$ instead of $\eta_{abcde}$ itself.

\subsubsection{$\GL{10}$ diffeomorphisms and Killing vectors} \label{s:DiffeosAndKilling}
We know that $\GL{10}$ diffeomorphisms act on tangent vectors via the usual Lie derivative
\begin{equation}
 L_U V^{M} = U^{N} \partial_{N} V^{M} - V^{N} \partial_{N} U^{M} \,.
\end{equation}
How does a $\GL{10}$ diffeomorphism act on a fundamental vector? A general diffeomorphism will not preserve the $\G$-structure and thus not act as an automorphism of $E_5$. In order to obtain an action on $E_5$, we have to restrict ourselves to automorphisms of the $\G$-structure, or infinitesimally to $\G$-Killing vectors.\footnote{We would like to thank Daniel Waldram for helpful discussions on this point.} These satisfy
\begin{equation}
 L_U \gE_{ab} = \frac12 \lambda_{ab}{}^{cd} \gE_{cd} \,,
\end{equation}
where $\lambda_{ab}{}^{cd} \in \mathfrak{gl(5)^+}$. Thus,
\begin{equation}
 \gE^{cd}{}_{M} L_U \gE_{ab}{}^{M} \in \mathfrak{gl(5)^+} \,.
\end{equation}
We can express this in terms of the projector onto the adjoint of $\G$, $\mathbb{P}_{adj}$, as
\begin{equation}
 \left( \mathbb{I} - \mathbb{P}_{adj} \right)_{ab,ef}^{cd,gh} \gE^{ef}{}_{M} L_U \gE_{gh}{}^{M} = 0 \,.
\end{equation}
The projector is explicitly given by
\begin{equation}
 \left(P_{adj}\right)^{ab,ef}_{cd,gh} = - \frac23 \delta^{ab}_{cd} \delta^{ef}_{gh} + \frac83 \delta^{ab}_{d[g} \delta^{ef}_{h]c} - \frac83 \delta^{ab}_{c[g} \delta^{ef}_{h]d} \,,
\end{equation}
Finally, it is useful to write this condition on Killing vectors $U^{M}$ in terms of an $\gE$-compatible connection $\nabla$ with $\GL{10}$ torsion $T_{MN}{}^{P}$.
\begin{equation}
 \left( \delta^{cd}_{ef} \delta^{gh}_{ab} - \frac14 \left(\mathbb{P}_{adj}\right)^{cd,gh}_{ab,ef} \right) \gE^{ef}{}_M \gE_{gh}{}^N \left(\nabla_N U^M + U^P T_{PN}{}^M \right) = 0 \,. \label{eq:KillingVectorP}
\end{equation}
This implies that a Killing vector $U^M$ must satisfy
\begin{equation}
 \begin{split}
  0 &= \frac14 \nabla_N U^M + \frac{1}{24} \delta^M_N \nabla_P U^P - \frac{1}{12} \gE_{dg}{}^M \gE^{cd}{}_N \nabla_{hc} U^{gh} \\
  & \quad + \frac14 U^L T_{LN}{}^M + \frac{1}{24} \delta^M_N U^L T_{LP}{}^P - \frac{1}{12} \gE_{dg}{}^M \gE^{cd}{}_N  U^L T_{LP}{}^Q \gE_{hc}{}^P \gE^{gh}{}_Q \,.
 \end{split} \label{eq:KillingVector}
\end{equation}
We can now define the Lie derivative of fundamental vector fields with respect to Killing vectors \eqref{eq:KillingVector}, in terms of the connection $\nabla$, as follows
\begin{dmath}
 L_U V^a = U^{M} \nabla_{M} V^a + \frac{1}{6} \left(\mathbb{P}_{adj}\right)_{b,ef}^{a,cd} V^b \gE_{cd}{}^M \gE^{ef}{}_N \nabla_M U^N - \frac{5}{24} V^a T_{MN}{}^M U^N + \gE_{bc}{}^P \gE^{ac}{}_N V^b T_{PM}{}^N U^M  \,. \label{eq:LieFun}
\end{dmath}
Here
\begin{equation}
 \left( P_{adj} \right)^{a,ef}_{b,cd} = \frac{1}{8} \delta^a_b \delta^{ef}_{cd} + \delta^a_{[c} \delta^{ef}_{d]b} \,.
\end{equation}
denotes the adjoint action on the fundamental representation of $\G$. The coefficient in front of $P_{adj}$ and the torsion terms are chosen in order for \eqref{eq:LieFun} to be independent of the choice of $\G$-connection.

Let us briefly expand on the connections appearing here. We require the connections in \eqref{eq:KillingVectorP}, \eqref{eq:KillingVector} and \eqref{eq:LieFun} to be compatible with $\gE_{ab}$, i.e. they induce a ``spin connection'' $\omega_{M,a}{}^b$ in order to satisfy the Vielbein principle
\begin{equation}
 \nabla_M \gE_{ab}{}^N = \partial_M \gE_{ab}{}^N + \Gamma_{MP}{}^N \gE_{ab}{}^P + 2 \omega_{M,[a}{}^c \gE_{b]c}{}^N = 0\,.
\end{equation}
We have a $\G$-connection when
\begin{equation}
 \nabla_M \eta_{abcde} = \partial_M \eta_{abcde} - \frac15 \eta_{abcde} \omega_{M,f}{}^f = 0 \,,
\end{equation}
This fixes
\begin{equation}
 \omega_{M,a}{}^a = \frac12 \partial_M \ln |\gE| \,,
\end{equation}
so that $\omega_{M,a}{}^b$ is a $\mathfrak{gl(5)}^+$-valued one-form. When $\omega_{M,a}{}^a = 0$ we have a $\SLf$-connection, which is what we will make use of in section \ref{s:Connection}.

\subsection{Generalised Lie derivative and closure constraints}
Similar to $\DWZW$ \cite{Blumenhagen:2014gva,Blumenhagen:2015zma,Bosque:2015jda}, we now define the exceptional field theory on the flattened spaces associated with the $\GL{10}$ vielbeine $\gE_{ab}{}^{M}$.\footnote{Despite this similarity our formulation is crucially different from $\DWZW$ because we do not use the Maurer-Cartan forms of the background gauge group.} Thus, we make use of the anholonomic derivatives
\begin{equation}
 D_{ab} \equiv \gE_{ab} = \gE_{ab}{}^{M} \partial_{M} \,,
\end{equation}
and define the generalised Lie derivative acting on $V \in \Gamma\left(E_5\right)$ as
\begin{equation}
 \gL_\Lambda V^{a} = \frac{1}{2} \Lambda^{bc} D_{bc} V^a + \frac{1}{5} V^a D_{bc} \Lambda^{bc} - V^b D_{bc} \Lambda^{ac} + \frac{1}{2} \tau_{bc,d}{}^a \Lambda^{bc} V^d \,. \label{eq:GenLieDeriv}
\end{equation}
For now we only require $\tau_{bc,d}{}^a$ to be constant but otherwise arbitrary, and will determine it soon. The generalised Lie derivative generates a $\SLf$ action if $\tau_{ab,c}{}^c = 0$. Otherwise, $\tau_{ab,c}{}^c$ generates an additional $\mathbb{R}^+$ action. In EFT, weighted vectors play an important role. For a vector of weight $w$, we thus define the generalised Lie derivative
\begin{equation}
 \gL_\Lambda V^{a} = \frac{1}{2} \Lambda^{bc} D_{bc} V^a + \left(\frac{1}{5} + \frac{w}{2}\right) V^a D_{bc} \Lambda^{bc} - V^b D_{bc} \Lambda^{ac} + \frac{1}{2} \tau_{bc,d}{}^a \Lambda^{bc} V^d + w\, \tau_{bc} \Lambda^{bc} V^a \,. \label{eq:GenLieDerivWeight}
\end{equation}
This mirrors the form of the genersalised Lie derivative of gauged EFTs \cite{Berman:2012uy}.\footnote{While this paper was being prepared for submission, we became aware of \cite{Ciceri:2016dmd} which considers deformations similar to \eqref{eq:GenLieDerivWeight} in the context of an EFT for massive IIA theory.}

As in EFT, we take the parameters of generalised diffeomorphisms $\Lambda$ to have weight $w = \frac15$ under the generalised Lie derivative and ask for the algebra of these generalised Lie derivatives to close, i.e.
\begin{equation}
 \left[ \gL_{\Lambda_1} ,\, \gL_{\Lambda_2} \right] = \gL_{\left[\Lambda_1, \, \Lambda_2\right]_E} \,, \label{eq:Closure}
\end{equation}
where
\begin{equation}
 \left[ \Lambda_1 , \, \Lambda_2 \right]_E = \frac12 \left( \gL_{\Lambda_1} \Lambda_2 - \gL_{\Lambda_2} \Lambda_1 \right) \,.
\end{equation}
This ensures covariance of the expression \eqref{eq:GenLieDeriv}. A  straightforward but tedious calculation, detailed in appendix \ref{A:Closure} shows that this is achieved when we impose four types of constraints, which we will collectively refer to as the closure constraints. First, we must identify $\tau_{ab,c}{}^d$ in \eqref{eq:GenLieDeriv} with the coefficients of anholonomy as follows
\begin{equation}
 \left[ D_{ab}, \, D_{cd} \right] = - 2 \tau_{ab,[c}{}^e D_{d]e} + \frac25 \tau_{ab} D_{cd} \,. \label{eq:AnholCoeff}
\end{equation}
In addition, we find that $\tau_{ab,c}{}^d$ must satisfy the linear and quadratic constraints of seven-dimensional maximal gauged supergravities and thus we interpret $\tau_{ab,c}{}^d$ as the \emph{background} embedding tensor, corresponding to the background vielbeine $\gE_{ab}{}^M$. The linear constraint restricts $\tau_{ab,c}{}^d$ to lie in the $\mathbf{15} \oplus \mathbf{40}' \oplus \mathbf{10}$ of $\SL{5}$ and relates it to $\tau_{ab}$. In particular, it has to satisfy
\begin{equation}
 \tau_{ab,c}{}^c = 0 \,, \qquad \tau_{c[a,b]}{}^c = \frac65 \tau_{ab} \,.
\end{equation}
Thus, we can write
\begin{equation}
 \tau_{ab,c}{}^d = \frac12 \delta_{[a}^d S_{b]c} + \frac12 \epsilon_{abcef} Z^{ef,d} + \frac{2}{15} \delta_c^d \tau_{ab} + \frac23 \delta^d_{[a} \tau_{b]c} \,, \label{eq:LCTau}
\end{equation}
where $Z^{(ab),c} = Z^{[ab,c]} = 0$, $S_{[ab]} = 0$ and $\tau_{(ab)} = 0$. Note that the embedding tensor here is related to the one in  \cite{Blair:2014zba}, $\hat{\tau}_{ab,c}{}^d$, by
\begin{equation}
 \hat{\tau}_{ab,c}{}^d = \tau_{ab,c}{}^d + \frac15 \delta^d_c \tau_{ab} \,, \qquad \hat{\tau}_{ab,c}{}^d = \frac12 \delta^d_{[a} S_{b]c} + \frac12 \epsilon_{abcef} Z^{ef,d} + \delta^d_{[c} \tau_{ab]} \,.
\end{equation}
The quadratic constraint can be written as
\begin{equation}
 2 \hat{\tau}_{ab,[c}{}^h \hat{\tau}_{d]h,e}{}^f - \hat{\tau}_{ab,e}{}^h \hat{\tau}_{cd,h}{}^f + \hat{\tau}_{ab,h}{}^f \hat{\tau}_{cd,e}{}^h = 0 \,. \label{eq:QC}
\end{equation}
Together with the linear constraints this leads to the expressions \cite{Malek:2015hma}
\begin{equation}
 \begin{split}
  \frac14 S_{ad} Z^{d(b,c)} - \frac14\, \epsilon_{adefg} Z^{de,b}Z^{fg,c} + \frac13\,\tau_{ad} Z^{d(b,c)} &=-\frac19\, \delta_{a}^{(b}\, \epsilon^{c)defg} \tau_{de}\tau_{fg} \,, \\
  S_{ad}Z^{bc,d} +\frac{1}{6}\,\epsilon^{bcdef}\,\tau_{ef} S_{ad} &= -\, \delta_{a}^{[b}\, \epsilon^{c]defg} \tau_{de}\tau_{fg} \,, \\
  \frac14 S_{ad}Z^{bc,d} +\frac13\,\tau_{ad}Z^{bc,d} &= -\frac29\, \delta_{a}^{[b}\, \epsilon^{c]defg} \tau_{de}\tau_{fg} \,.
 \end{split} \label{eq:QCLin}
\end{equation}
Finally, we require a ``section condition'' for the anholonomic derivatives
\begin{equation}
 \begin{split}
  D_{[ab} \otimes D_{cd]} &= 0 \,, \\
  D_{[ab} D_{cd]} + 2 \tau_{[ab} D_{cd]} &= 0 \,, \label{eq:SecCon}
 \end{split}
\end{equation}
where the $\otimes$ in the first line denotes that the derivatives act on two different objects. Note that the symmetric part of \eqref{eq:AnholCoeff} together with the linear constraint \eqref{eq:LCTau} implies that
\begin{equation}
 \left( Z^{ab,c} - \frac13 \epsilon^{abcde} \tau_{de} \right) D_{ab} = 0 \,. \label{eq:IntertwineConstraint}
\end{equation}
At this stage we would once again like to emphasise the difference to $\DWZW$. There, the background vielbein would be described by the Maurer-Cartan form of the gauge group, which by the above constraint \eqref{eq:IntertwineConstraint} can have less than 10 dimensions. Thus, if we had wanted to use the Maurer-Cartan form here the extended manifold would have to have less than 10 dimensions. A further discussion on this subject will appear in \cite{TogetherPaper}.

Finally, we can use expression \eqref{eq:AnholCoeff} to determine the different irreps of the background embedding tensor in terms of the vielbeine $\gE_{ab}{}^{M}$. We find
\begin{equation}
 \begin{split}
  \tau_{ab} &= - \frac13 \left( \partial_M \gE_{ab}{}^M + D_{ab} \ln \gE \right) \,, \\
  S_{ab} &= \frac23 \gE^{ef}{}_M D_{e(a} \gE_{b)f}{}^M \,, \\
  Z^{ab,c} &= - \frac{1}{15} \epsilon^{abdef} \gE^{cg}{}_M \left(D_{fg} \gE_{de}{}^M - D_{de} \gE_{fg}{}^M \right) + \frac{2}{45} \epsilon^{abcde} \left( \partial_M \gE_{de}{}^M + D_{de} \ln\gE \right) \,. \label{eq:EmbTensorBack}
 \end{split}
\end{equation}
Here $\gE$ denotes the determinant of $\gE^{ab}{}_M$. In order to satisfy the linear constraint, we also have to impose that the following vanishes:
\begin{equation}
 \begin{split}
  0 &= \epsilon^{bcfgh} \gE^{de}{}_M \left(D_{ha} \gE_{fg}{}^M - D_{fg} \gE_{ha}{}^M \right) \\
  & \quad - \frac13 \epsilon^{bcdef} \left( D_{af} \ln \gE - \gE^{gh}{}_M D_{ag} \gE_{hf}{}^M + \partial_M \gE_{af}{}^M - \gE^{gh}{}_M D_{fg} \gE_{hf}{}^M \right) \,. \label{eq:LCBack}
 \end{split}
\end{equation}

\subsubsection{Comparison to standard and gauged EFT}
Let us reflect and compare the situation here to the usual  formulation of exceptional field theory. This discussion is very similar to that in $\DWZW$, see \cite{Blumenhagen:2014gva,Blumenhagen:2015zma,Bosque:2015jda}, although our vielbeine are not necessarily Maurer-Cartan forms. Our 10-manifold has a $\G$-structure, which, when it is not locally flat, introduces a manifest background dependence through the vielbeine $\gE_{ab}$. This is captured by the coefficients of anholonomy of the derivatives \eqref{eq:AnholCoeff}, introduces a gauging in the generalised Lie derivative \eqref{eq:GenLieDeriv} and is identified with the background embedding tensor. Closure of the algebra of generalised diffeomorphisms further requires a ``section condition'' \eqref{eq:SecCon}. The theory thus resembles an expansion around an EFT background, as in the ``gauged EFT'' setup \cite{Berman:2012uy,Blair:2014zba}.

In the gauged EFT setup, just as in gauged DFT, the embedding tensor is determined in terms of some $\G$ ``twist matrices''
\begin{equation}
 W_{ab}{}^{ij} = \rho^{-1} U_{[a}{}^i U_{b]}{}^j\,, \label{eq:WTwist}
\end{equation}
where $|U| = 1$ and $\rho$ is a scalar density. The precise relationship is given by the generalised Lie derivative of EFT
\begin{equation}
 \gL^{0}_{W_{ab}} W_{cd} = \frac12 \tau_{ab,cd}{}^{ef} W_{ef} \,, \label{eq:gEmbeddingTensor}
\end{equation}
$\gL^0$ here has the same form as \eqref{eq:GenLieDerivWeight} but with $\tau_{ab,c}{}^d = \tau_{ab} = 0$, $W_{ab}$ has weight $w = \frac15$ and
\begin{equation}
 \tau_{ab,cd}{}^{ef} = 4 \tau_{ab,[c}{}^{[e} \delta_{d]}^{f]} + \frac45 \tau_{ab} \delta_{cd}^{ef} \,. \label{eq:BigTau}
\end{equation}
In terms of the irreducible representations this gives
\begin{equation}
\begin{split}
 S_{ab} &= - \frac1\rho \partial_{ij} U_{(a}{}^i U_{b)}{}^{j} \,,\\
 Z^{ab,c} &= \frac{1}{2\rho} \epsilon^{ijklm} \left( U_{lm}{}^{ab}  \, \partial_{ij} U_k{}^{c} - U_{lm}{}^{[ab}\,\partial_{ij} U_k{}^{c]} \right) \,, \\
 \tau_{ab} &= - \frac{1}{2\rho} \partial_{ij} U_{ab}{}^{ij} - 6\,\rho^{-1}  \, U_{ab}{}^{ij} \,\partial_{ij} \rho \,.
 \label{consistent}
\end{split} 
\end{equation}

We see that \eqref{eq:AnholCoeff}, which can be rewritten in the more suggestive form
\begin{equation}
 L_{\gE_{ab}} \gE_{cd} = \frac12 \tau_{ab,cd}{}^{ef} \gE_{ef} \,, \label{eq:LieE}
\end{equation}
is similar in spirit, but there the background vielbeine $\gE_{ab}$ do not have to be $\G$-valued, and we use the conventional Lie derivative, not the generalised Lie derivative. Nonetheless, in gauged EFT one also finds that the section condition can be relaxed, see for example the analogous discussion for gauged DFT \cite{Aldazabal:2011nj,Grana:2012rr,Geissbuhler:2013uka,Berman:2013uda} and also the review \cite{Aldazabal:2013sca}: for closure of the algebra one must impose the quadratic constraint on the embedding tensor, which by \eqref{eq:gEmbeddingTensor} automatically satisfies the linear constraint, and the section condition \eqref{eq:SecCon}, where in this case the background vielbeine would be $W_{ab}$, \eqref{eq:WTwist}. However, one also imposes the section condition between the background and fluctuations,
\begin{equation}
 \partial_{[ij} W_{|ab|}{}^{kl} \partial_{mn]} = 0 \,, \label{eq:SCBackFluct}
\end{equation}
when acting on any fluctuations.

Thus, the curved EFT formulation looks similar to gauged EFT when we have a locally flat $\G$-structure, as in eq. \eqref{eq:CanStructure}. However, even in the locally flat case, there is the difference that the embedding tensor would still be given by \eqref{eq:AnholCoeff} rather than \eqref{eq:gEmbeddingTensor}, and that we do not need to impose \eqref{eq:SCBackFluct}. A straightforward calculation shows, however, that in the locally flat case, where we can write
\begin{equation}
 \gE_{ab}{}^{ij} = \rho^{-1} U_{[a}{}^i U_{b]}{}^j \,,
\end{equation}
\eqref{eq:EmbTensorBack} agrees with \eqref{eq:gEmbeddingTensor}. Furthermore, one finds that in this case \eqref{eq:SCBackFluct} is sufficient to satisfy \eqref{eq:IntertwineConstraint}, i.e.
\begin{equation}
 \left( Z^{ab,c} - \frac13 \epsilon^{abcde} \tau_{de} \right) D_{ab} = 0 \,,
\end{equation}
since this is always taken to act on fluctuations. Finally, it is easy to see that when we impose \eqref{eq:SCBackFluct}, \eqref{eq:LieE} and \eqref{eq:gEmbeddingTensor} agree. This implies that \eqref{eq:SCBackFluct} and local flatness are sufficient for the the vielbeine to satisfy the linear constraint, i.e. \eqref{eq:LCBack}. To summarise, in the locally flat case with \eqref{eq:SCBackFluct} our formalism reduces to the usual EFT set-up.

In the following sections we will show that even when the $\G$-structure is not locally flat and we do not impose \eqref{eq:SCBackFluct}, we can use the $\GL{10}$ vielbeine $\gE_{ab}$ to construct a curved EFT formulation reminiscent of gauged EFT. However, the fact that the ``background'' is described by a $\GL{10}$ object while the fluctuations are in $\G$ means that the theory is not background-independent, see for example the discussion in section 5 of \cite{Hohm:2015ugy}. In contrast the usual double field theory formulation, which we wish to interpret as the ``locally flat'' case, has recently been confirmed to be background-independent \cite{Hohm:2015ugy} and it is reasonable to expect the same to be true of exceptional field theory.

Nonetheless, the formulation presented here is manifestly coordinate invariant, and has a clear patching prescription which does not require the section condition. It can thus describe non-geometric backgrounds \cite{Cederwall:2014opa}. The interested reader can find a review of the patching discussion in double field theory in \cite{Hohm:2013bwa}. Finally, one may hope that it captures other effects, such as non-Abelian T-duality \cite{delaOssa:1992vci}.

\subsection{$\GL{10}$ covariant derivative} \label{s:Connection}
Following \cite{Blumenhagen:2014gva}, we define a spin-connection for a vector $V$
\begin{equation}
 \nabla_{ab} V^c = D_{ab} V^c + \Conn_{ab,d}{}^c V^d \,,
\end{equation}
such that we can rewrite the generalised Lie derivative \eqref{eq:GenLieDerivWeight} as
\begin{equation}
 \gL_\Lambda V^a = \frac{1}{2} \Lambda^{bc} \nabla_{bc} V^a + \left(\frac{1}{5} + \frac{w}{2}\right) V^a \nabla_{bc} \Lambda^{bc} - V^b \nabla_{bc} \Lambda^{ac} \,. \label{eq:ConnLambda}
\end{equation}
We then find
\begin{equation}
 \Conn_{ab,c}{}^d = \frac16 \epsilon_{abcef} Z^{ef,d} + \frac18 \delta_{[a}^{d} S_{b]c} - \frac19 \delta_c^d \tau_{ab} - \frac59 \delta^d_{[a} \tau_{b]c} \,. \label{eq:Connection}
\end{equation}
Note that this is traceless
\begin{equation}
 \Conn_{ab,c}{}^c = 0 \,,
\end{equation}
and that for a scalar of weight $w$, we have
\begin{equation}
 \gL_\Lambda S = \frac12 \Lambda^{ab} D_{ab} S + \frac{w}{2} S \,D_{ab} \Lambda^{ab} = \frac12 \Lambda^{ab} \nabla_{ab} S + \frac{w}{2} S\, \nabla_{ab} \Lambda^{ab} \,,
\end{equation}
so that there is no ambiguity as to whether we should be using $\nabla_{ab}$ or $D_{ab}$ for the weight-term.
It is easy to check that
\begin{equation}
 \nabla_{ab} \epsilon_{cdefg} = \Conn_{ab,h}{}^h \epsilon_{cdefg} = 0\,,
\end{equation}
since $\left(\Conn_{ab}\right)$ is $\mathfrak{sl}(5)$-valued. Thus,
\begin{equation}
 \Conn_{M,a}{}^b = \frac12 \gE^{cd}{}_M \Conn_{cd,a}{}^b \,,
\end{equation}
is a $\mathfrak{sl}(5)$-valued one-form and it induces a connection for $\GL{10}$-diffeomorphisms, $\Gamma_{MN}{}^{P}$, via the vielbein postulate
\begin{equation}
 \nabla_{M} \gE_{ab}{}^{N} \equiv \partial_{M} \gE_{ab}{}^{N} + 2 \gE^{cd}{}_M \Conn_{cd,[a}{}^{f} \gE_{b]f}{}^{N}  + \Gamma_{MP}{}^{N} \gE_{ab}{}^{P} = 0 \,. \label{eq:VielbeinPostulate}
\end{equation}
It is easy to check that $\nabla_{M}$ defined in \eqref{eq:VielbeinPostulate} is a connection if $\Conn_{ab,c}{}^{d}$ is a $\GL{10}$-scalar. This follows from the tensorial definitions \eqref{eq:AnholCoeff}. As discussed in subsection \ref{s:DiffeosAndKilling}, the connection here is a $\SLf$-connection. 

Finally, using \eqref{eq:Connection} we obtain the explicit expression for the components of the $\GL{10}$-connection
\begin{equation}
 \begin{split}
  \Gamma_{MN}{}^{P} &= \gE^{ab}{}_N \left( - \frac12 \partial_M \gE_{ab}{}^P - \frac18 \gE_{bc}{}^P \gE^{cd}{}_M S_{ad} - \frac16 \gE^{cd}{}_M \gE_{bf}{}^P \epsilon_{acdgh} Z^{gh,f} - \frac59 \gE^{cd}{}_M \gE_{bc}{}^P \tau_{ad} \right) \\
  & \quad - \frac29  \delta_N^P \gE^{cd}{}_M \tau_{cd}  \,. \label{eq:BigConnection}
 \end{split}
\end{equation}

\subsubsection{Curvature, torsion and integration by parts}
Let us calculate the usual $\GL{10}$ curvature and torsion of this connection. The curvature is best calculated in terms of the spin connection $\omega_{M,a}{}^b$. It is given by
\begin{equation}
 R_{MN,a}{}^b = 2 \partial_{[M} \omega_{N],a}{}^b + 2 \omega_{[M,|a}{}^c \omega_{N],c}{}^b \,.
\end{equation}
One can check that this is still traceless so that $\left(R_{MN}\right)_a{}^b = R_{MN,a}{}^b$ is a $\mathfrak{sl}(5)$ element. Using the vielbeine $E_{ab}{}^M$ we see that the curvature tensor lives in the
\begin{equation}
 \mathbf{45} \otimes \mathbf{24} = \mathbf{5} \oplus \mathbf{45} \oplus \mathbf{45} \oplus \mathbf{50} \oplus \mathbf{70} \oplus \mathbf{105} \oplus \mathbf{280} \oplus \mathbf{480} \,. \label{eq:CurvatureIrreps}
\end{equation}
To evaluate the curvature tensor it helps to note that
\begin{equation}
 2 \partial_{[M} \omega_{N],a}{}^b = \partial_{[M} E^{cd}{}_{N]} \omega_{cd,a}{}^b \,,
\end{equation}
and
\begin{equation}
 \frac14 E^{ab}{}_M E^{cd}{}_N \left[ D_{ab}, D_{cd} \right] = \partial_{[N} E^{ab}{}_{M]} D_{ab} \,.
\end{equation}
But from \eqref{eq:AnholCoeff} and \eqref{eq:BigTau} we have that
\begin{equation}
 \partial_{[M} E^{ab}{}_{N]} = -\frac18 E^{cd}{}_{[M} E^{ef}{}_{N]} \tau_{cd,ef}{}^{ab} \,.
\end{equation}
The rest is a tedious but straightforward calculation which shows that none of the irreducible representations \eqref{eq:CurvatureIrreps} vanish, even using the quadratic constraints. We summarise the irreducible representations in appendix \ref{A:Curvature}.

The torsion of the connection is given by
\begin{equation}
 \begin{split}
  T_{MN}{}^P &= \Gamma_{MN}{}^P - \Gamma_{NM}{}^{P} = \gE^{ab}{}_{[M} \partial_{N]} \gE_{ab}{}^P + \frac14 \gE_{bc}{}^P \gE^{ab}{}_M \gE^{cd}{}_N S_{ad} \\
  & \quad - \frac13 \gE^{ab}{}_{[N} \gE^{cd}{}_{M]} \gE_{bf}{}^P \epsilon_{acdgh} Z^{gh,f} + \frac49 \delta_{[M}^P \gE^{cd}{}_{N]} \tau_{cd} \,.
 \end{split}
\end{equation}
We see that for a general background, the torsion of this connection does not vanish. Let us consider its trace
\begin{equation}
 T_{MN}{}^N = - 2 \gE^{cd}{}_{M} \tau_{cd} + \partial_M \ln\gE + \frac12 \gE^{ab}{}_M \partial_N \gE_{ab}{}^N = - \frac72 \gE^{ab}{}_M \tau_{ab} \,,
\end{equation}
where in the final step we used the relation \eqref{eq:EmbTensorBack}. This is important since it measures the obstruction to integrating by parts: when integrating by parts we will pick up terms such as
\begin{equation}
 I = \int d^{10}x\, \nabla_{M} V^M \,,
\end{equation}
where $V^M$ will be a diffeomorphism-density. Thus,
\begin{equation}
 I = \int d^{10}x \left( \partial_M V^M + \Gamma_{MN}{}^M V^N - \Gamma_{NM}{}^N V^N \right) = \int d^{10}x\, \left( \partial_M V^M + \frac72 \gE^{ab}{}_M \tau_{ab} V^M \right) \,.
\end{equation}
To integrate by parts we require $I$ to be a boundary term, which only occurs when $\tau_{ab} = 0$. This is consistent with the fact that supergravities with a trombone gauging do not admit an action principle. Instead they are defined only at the level of the equations of motion \cite{LeDiffon:2008sh}. From the gauged supergravity perspective, this makes sense because the trombone gauges an on-shell symmetry. Indeed, in the usual gauged EFT formulation, one also finds that the trombone is the obstruction to integration by parts by a similar argument to that presented here \cite{Hohm:2014qga}.

To conclude this section, let us note that there are trivial gauge parameters, with respect to which the Lie derivative vanishes. These are given by
\begin{equation}
 \Lambda_{triv}^{ab} = \epsilon^{abcde} \nabla_{cd} B_e \,,
\end{equation}
where $B_a$ is any element of $E^*_5$ of weight $\frac25$. This is the generalisation of an ``exact form'' as given by the generalised Cartan Calculus \cite{Hohm:2015xna,Wang:2015hca} that we will discuss in the following section.

\section{Tensor Hierarchy}\label{s:TH}
In the full EFT, the fields which are ``off-diagonal'' between the internal extended space and the external seven-dimensional space are described by a hierarchy of tensor fields. These are related to the tensor hierarchy of maximal gauged SUGRA \cite{deWit:2005hv,deWit:2008ta}. Their structure can be nicely described in terms of a certain chain complex \cite{Cederwall:2013naa,Hohm:2015xna,Wang:2015hca}. In section \ref{s:CCC} we first generalise the formulation of this chain complex \cite{Wang:2015hca} to take into account the curvature of the $\G$-structure. We then show in subsection \ref{s:TensorHier} how this can be used to describe the tensor hierarchy. Finally we derive the topological term of the Lagrangian in subsection \ref{s:TopTerm}.

\subsection{Curved Cartan Calculus} \label{s:CCC}
We begin by constructing the curved version of the generalised Cartan Calculus \cite{Hohm:2015xna,Wang:2015hca}. We want to introduce a nilpotent derivative so that we obtain a chain complex
\begin{equation}
 \TAwl \xleftarrow{~\hat{\partial}~} \TBwl \xleftarrow{~\hat{\partial}~} \TCwl \xleftarrow{~\hat{\partial}~} \TDwl \,, \label{eq:ChainComplex}
\end{equation}
between the modules required for the tensor hierarchy, summarised in table \ref{t:Forms}.

\vspace{1em}
\noindent\makebox[\textwidth]{
 \begin{minipage}{\textwidth}
  \begin{center}
  \begin{tabular}{|c|c|c|c|}
  \hline
   Module($w$,$\lambda$) & Representations & Gauge field & Field strength \Tstrut\Bstrut \\ \hline
   $\TAwl$ & $\mathbf{10}$ & $\Aa^{ab}$ & $\Fa^{ab}$ \\
   $\TBwl$ & $\overline{\mathbf{5}}$ & $\Ab_a$ & $\Fb_a$ \\
   $\TCwl$ & $\mathbf{5}$ & $\Ac^{a}$ & $\Fc^{a}$ \\
   $\TDwl$ & $\overline{\mathbf{10}}$ & $\Ad_{ab}$ & $\Fd_{ab}$ \\
   \hline
  \end{tabular}
 \vskip-0.5em
 \captionof{table}{\small{Modules, gauge fields and field strengths relevant for the tensor hierarchy and their representations under $\SL{5}$ and $\GL{10}$. $w$ denotes their weight under generalised Lie derivatives while $\lambda$ denotes their weight under $\GL{10}$ diffeomorphisms.}} 
 \label{t:Forms}
  \end{center}
 \end{minipage}
}
\vspace{1em}

\noindent In order to avoid clutter we will from here onwards drop the $\lambda$ value when referring to the modules in table \ref{t:Forms}, with the $\GL{10}$-values always to be taken as in table \ref{t:Forms}. We will also make use of a scalar density $\TSwl$ which has weight $1$ under both the generalised Lie derivative and $\GL{10}$-diffeomorphisms, but again we will refer to it simply as $\TSw$. We also define a bilinear product $\bullet$ between certain modules, which maps as follows.

\vspace{1em}
\noindent\makebox[\textwidth]{
 \begin{minipage}{\textwidth}
  \begin{center}
  \begin{tabular}{c| c c c c}
   $\bullet$ & $\TAw$ & $\TBw$ & $\TCw$ & $\TDw$ \\
   \hline
   $\TAw$ & $\TBw$ & $\TCw$ & $\TDw$ & $\TSw$ \\
   $\TBw$ & $\TCw$ & $\TDw$ & $\TSw$ & \\
   $\TCw$ & $\TDw$ & $\TSw$ & & \\
   $\TDw$ & $\TSw$ & & & 
  \end{tabular}
  \end{center}
 \end{minipage}
} \\

\noindent Finally, we want the nilpotent derivative $\hat{\partial}$ and the product $\bullet$ to obey the following identity \cite{Hohm:2015xna,Wang:2015hca}: for all $\Lambda \in \TAw$ and $ {\cal T} \in \TBw$ or $\TCw$,
\begin{equation}
 {\cal L}_\Lambda {\cal T} = \Lambda \bullet \left( \hat{\partial} {\cal T} \right) + \hat{\partial} \left( \Lambda \bullet {\cal T} \right) \,.
\end{equation}
We use the same $\bullet$ product as in the ``flat case'' \cite{Wang:2015hca}, defined as
\begin{equation}
 \begin{split}
  \left(\TA_1 \bullet \TA_2\right)_a &= \frac14 \epsilon_{abcde} \TA_1^{bc} \TA_2^{de} \,, \\
  \left(\TA \bullet \TB\right)^a &= \TA^{ab} \TB_{b} \,, \\
  \left(\TA \bullet \TC\right)_{ab} &= \frac14 \epsilon_{abcde} \TA^{cd} \TC^e \,, \\
  \TA \bullet \TD &= \frac12 \TA^{ab} \TD_{ab} \,, \\
  \left( \TB_1 \bullet \TB_2 \right)_{ab} &= \TB_{2[a} \TB_{|1|b]} \,, \\
  \TB \bullet \TC &= \TB_a \TC^a \,, \label{eq:Bullet}
 \end{split}
\end{equation}
and which is defined to be symmetric when acting on different modules. However, we modify the derivative $\hat{\partial}$ to be
\begin{equation}
 \hat{\partial} \TB^{ab} = \frac{1}{2} \epsilon^{abcde} \nabla_{cd} \TB_e \,, \qquad \hat{\partial} \TC_a = \nabla_{ba} \TC^b \,, \qquad \hat{\partial} \TD^a = \frac12 \epsilon^{abcde} \nabla_{bc} \TD_{de} \,, \label{eq:NilPot}
\end{equation}
where $\TB \in \TBw$, $\TC \in \TCw$ and $\TD \in \TDw$. Note that these definitions also map the $\GL{10}$ weights as required, see table \ref{t:Forms}.

The derivative $\nabla_{ab}$ is as in \eqref{eq:Connection} and it is important to note that $\hpartial$ thus satisfies integration by parts when $\tau_{ab} = 0$. Let us now check the nilpotency, starting with
\begin{equation}
 \left( \hat{\partial}\hat{\partial} \TC \right)^{ab} = \frac12 \epsilon^{abcde} \nabla_{cd} \nabla_{fe} \TC^f \,.
\end{equation}
We can split this expression into terms quadratic in the embedding tensor components, those linear in the embedding tensor components and those without. For those without we find
\begin{equation}
 \left(\hpartial\hpartial \TC \right)^{ab}_{0} = -\frac12 \epsilon^{abcde} D_{cd} D_{ef} \TC^f \,.
\end{equation}
We use the identity
\begin{equation}
 2 D_{[ab} D_{cd]} = 2 D_{a[b} D_{cd]} + \left[ D_{[cd} , D_{|a|b]} \right] \,,
\end{equation}
and the section condition \eqref{eq:SecCon} to write this as
\begin{equation}
 \left(\hpartial \hpartial \TC\right)^{ab}_{0} = - \frac14 \epsilon^{abcde} \left[ D_{ef} , D_{cd} \right] \TC^f \,.
\end{equation}
It is now easy to check using the coefficients of anholonomy \eqref{eq:AnholCoeff}, the linear constraint \eqref{eq:LCTau} and \eqref{eq:Connection} that the terms linear in $S_{ab}$, $Z^{ab,c}$ and $\tau_{ab}$ vanish. The terms quadratic in the embedding tensor vanish by the quadratic constraint \eqref{eq:QC}. The same steps can be used to show that
\begin{equation}
 \left(\hpartial \hpartial \TD\right)_a = 0 \,,
\end{equation}
thus showing that the derivative $\hpartial$ is nilpotent. One can also check that this nilpotent derivative $\hpartial$ is covariant under generalised Lie derivatives in the sense that the following diagram commutes:
\begin{center}
\begin{tikzcd}[column sep=large, row sep=large]
 \TA \arrow{d}{\gL} & \TB \arrow{l}[swap]{\hpartial} \arrow{d}{\gL} & \TC \arrow{l}[swap]{\hpartial} \arrow{d}{\gL} & \TD \arrow{l}[swap]{\hpartial} \arrow{d}{\gL} \\
 \TA & \TB \arrow{l}[swap]{\hpartial} & \TC \arrow{l}[swap]{\hpartial} & \TD \arrow{l}[swap]{\hpartial}
\end{tikzcd}
\end{center}

\subsection{Tensor Hierarchy} \label{s:TensorHier}
We now construct the tensor hierarchy \cite{deWit:2005hv,deWit:2008ta} as in EFT \cite{Hohm:2013vpa} by introducing field strengths of the various potentials in table \ref{t:Forms}. \emph{Mutatis mutandis}, the construction in this section is formally identical to that presented in \cite{Wang:2015hca}. That is, the arguments and formulae in \cite{Hohm:2015xna,Wang:2015hca} hold, subject to the modification of the generalised Lie derivative \eqref{eq:GenLieDeriv} and the nilpotent derivative \eqref{eq:NilPot}. Thus, we will keep the discussion here brief and refer the interested readers to the original construction in $E_6$ \cite{Hohm:2013vpa} as well as \cite{Hohm:2015xna,Wang:2015hca}.

The fields of the tensor hierarchy are forms of the external spacetime as well as forms of the extended space, i.e. of the chain complex \eqref{eq:ChainComplex}. Because they can depend on both the external spacetime and the extended space, they will transform under generalised diffeomorphisms, $\GL{10}$-diffeomorphisms and external diffeomorphisms. To account for these different symmetries, we introduce a covariant derivative for the external directions \cite{Hohm:2013vpa}
\begin{equation}
 \D_\mu = \partial_\mu - {\cal L}_{\Aa_\mu} \,.
\end{equation}
Its commutator defines a field strength
\begin{equation}
 \left[ \D_\mu, \, \D_\nu \right] = -\gL_{F_{\mu\nu}} \,, \label{eq:DComm}
\end{equation}
where
\begin{equation}
 F_{\mu\nu} = 2 \partial_{[\mu} \Aa_{\nu]} - \left[ \Aa_{\mu} ,\, \Aa_\nu \right]_E \,. \label{eq:NaiveF}
\end{equation}
Here $\left[V , \, W\right]_E = \frac12 \left( \gL_{V} W - \gL_W V \right)$ is the antisymmetrisation of the generalised Lie derivative. Although \eqref{eq:DComm} is manifestly invariant under generalised Lie derivatives, the naive field strength $F_{\mu\nu}$ as defined in \eqref{eq:NaiveF} is not. The deviation from covariance is however a term that generates a trivial generalised Lie derivative, i.e. it is of the form $\left(\hpartial \Ab_{\mu\nu}\right)^{ab}$. This intertwining between forms of different degrees is a defining feature of the tensor hierarchy, which continues by defining a field strength for $\Ab_{\mu\nu,a}$ etc.

Subject to the modifications of the generalised Lie derivative and the nilpotent operator $\hpartial$, we can proceed with formally equivalent definitions as for the ``flat'' case \cite{Wang:2015hca}. In particular, we define the covariant field strengths (we now drop the $\SLf$ indices to avoid clutter)
\begin{equation}
\begin{split}
\Fa_{\mu\nu} &= 2\partial_{[\mu}\Aa_{\nu]} - [\Aa_{\mu},\Aa_{\nu}]_E + \hpartial \Ab_{\mu\nu} \,, \\
\Fb_{\mu\nu\rho} &= 3\D_{[\mu}\Ab_{\nu\rho]} - 3\partial_{[\mu}\TA_{\nu}\bullet\TA_{\rho]} + \TA_{[\mu}\bullet[\TA_{\nu},\TA_{\rho]}]_E + \hpartial\Ac_{\mu\nu\rho} \,, \\
\Fc_{\mu\nu\rho\sigma} &= 4\D_{[\mu}\Ac_{\nu\rho\sigma]} + 3\hpartial\Ab_{[\mu\nu}\bullet\Ab_{\rho\sigma]} - 6\Fa_{[\mu}\bullet\Ab_{\nu\rho\sigma]} + 4\Aa_{[\mu}\bullet(\Aa_{\nu}\bullet\partial_{\rho}\Aa_{\sigma]}) \\
& \quad - \Aa_{[\mu}\bullet(\Aa_{\nu}\bullet[\Aa_{\rho},\Aa_{\sigma]}]_E) + \hpartial\Ad_{\mu\nu\rho\sigma} \,. \\
\end{split}
\label{eq:fieldstrengths}
\end{equation}
From these definitions, one can see that the field strengths satisfy the Bianchi identities
\begin{equation}
\begin{split}
 3\D_{[\mu}\Fa_{\nu\rho]} &= \hpartial\Fb_{\mu\nu\rho} \,, \\
 4\D_{[\mu}\Fb_{\nu\rho\sigma]} + 3\Fa_{[\mu\nu}\bullet\Fa_{\rho\sigma]} &= \hpartial\Fc_{\mu\nu\rho\sigma} \,.
\end{split}
\label{eq:Bianchi}
\end{equation}

Varying the gauge potentials leads to the following variations of the field strengths
\begin{equation}
 \begin{split}
  \delta\Fa_{\mu\nu} &= 2\D_{[\mu}\delta\TA_{\nu]} + \hpartial\Delta\TB_{\mu\nu} \,, \\
  \delta\Fb_{\mu\nu\rho} &= 3\D_{[\mu}\Delta\TB_{\nu\rho]} - 3\delta\TA_{[\mu}\bullet\Fa_{\nu\rho]} + \hpartial\Delta\TC_{\mu\nu\rho} \,, \\
  \delta\Fc_{\mu\nu\rho\sigma} &= 4\D_{[\mu}\Delta\TC_{\nu\rho\sigma]} - 4\delta\TA_{[\mu}\bullet\Fb_{\nu\rho\sigma]} - 6\Fa_{[\mu}\bullet\Delta\TB_{\nu\rho\sigma]} + \hpartial\Delta\TD_{\mu\nu\rho\sigma} \,,
 \end{split}
\label{eq:fieldstrengthvariation}
\end{equation}
where we defined the ``covariant'' gauge field variations
\begin{equation}
\begin{split}
 \Delta\TB_{\mu\nu} &= \delta\TB_{\mu\nu} + \TA_{[\mu}\bullet\delta\TA_{\nu]} \,, \\
 \Delta\TC_{\mu\nu\rho} &= \delta\TC_{\mu\nu\rho} - 3\delta\TA_{[\mu}\bullet\TB_{\nu\rho]} + \TA_{[\mu}\bullet(\TA_{\nu}\bullet\delta\TA_{\rho]}) \,, \\
 \Delta\TD_{\mu\nu\rho\sigma} &= \delta\TD_{\mu\nu\rho\sigma} - 4\delta\TA_{[\mu}\bullet\TC_{\nu\rho\sigma]} + 3\TB_{[\mu\nu}\bullet(\delta\TB_{\rho\sigma]}+2\TA_{\rho}\bullet\delta\TA_{\sigma]}) + \TA_{[\mu}\bullet(\TA_{\nu}\bullet(\TA_{\rho}\bullet\delta\TA_{\sigma]})) \,.
\end{split}
\label{eq:gaugefieldvariation}
\end{equation}
Finally, the field strengths are invariant under the gauge transformations given by
\begin{equation}
 \begin{split}
  \delta A_{\mu} &= D_{\mu} \Lambda - \hat{\partial} \Xi_\mu \,, \\
  \Delta B_{\mu\nu} &= \Lambda \bullet {\cal F}_{\mu\nu} + 2 D_{[\mu} \Xi_{\nu]} - \hat{\partial} \Theta_{\mu\nu} \,, \\
  \Delta C_{\mu\nu\rho} &= \Lambda \bullet {\cal H}_{\mu\nu\rho} + 3 {\cal F}_{[\mu\nu} \bullet \Xi_{\rho]} + 3 D_{[\mu} \Theta_{\nu\rho]} - \hat{\partial} \Omega _{\mu\nu\rho} \,, \\
  \Delta D_{\mu_\nu\rho\sigma} &= \Lambda \bullet {\cal J}_{\mu\nu\rho\sigma} - 4 {\cal H}_{[\mu\nu\rho} \bullet \Xi_{\sigma]} + 6 {\cal F}_{[\mu\nu} \bullet \Theta_{\rho\sigma]} + 4 D_{[\mu} \Omega_{\nu\rho\sigma]} \,. \label{eq:GaugeTransformations}
 \end{split}
\end{equation}

\subsection{Topological Term} \label{s:TopTerm}
We now wish to construct the analogue of the topological term of EFT \cite{Musaev:2015ces} which reduces in the locally flat case to the topological term of seven-dimensional maximal gauged SUGRA \cite{Samtleben:2005bp}. Using the formalism described above, we construct it as a boundary term in eight external and ten extended dimensions. The proposed term is
\begin{equation}
 S_{top} = - \frac{1}{2\sqrt{6}} \int \mathrm{d}^{10}Y\, \mathrm{d}^8x \left( \frac14 \hpartial \Fc_{\mu_1\ldots\mu_4} \bullet \Fc_{\mu_5\ldots\mu_8} - 4 \Fa_{\mu_1\mu_2} \bullet \left( \Fb_{\mu_3\ldots\mu_5} \bullet \Fb_{\mu_6\ldots\mu_8} \right) \right) \epsilon^{\mu_1\ldots\mu_8} \,. \label{eq:TopTerm}
\end{equation}
Here we abuse notation by labelling the eight-dimensional space and the seven-dimensional external space that is its boundary by the same indices, i.e. $\mu = 1, \ldots, 8$ above. It is easy to check that the integrand has the appropriate weight under generalised diffeomorphisms, $\GL{10}$-diffeomorphisms and external diffeomorphisms. We will show that when the trombone vanishes, the variation of \eqref{eq:TopTerm} is a boundary term, because this is sufficient for calculating the action. We use the fact that when $\tau_{ab} = 0$ we can integrate the nilpotent derivative $\hpartial$ by parts, to obtain

\begin{equation}
 \begin{split}
  \delta S_{top} &= -\frac{1}{2\sqrt{6}} \int d^{10}Y\, d^8x \left[ - 8 \D_{\mu_1} \left( \delta \TA_2 \bullet \left( \Fb_{\mu_3\mu_4\mu_5} \bullet \Fb_{\mu_6\mu_7\mu_8} \right) \right) + 2 \D_{\mu_1} \left( \hpartial \Delta \TC_{\mu_2\mu_3\mu_4} \bullet \Fc_{\mu_5\ldots\mu_8} \right) \right. \\
  & \quad \left. - 24 \D_{\mu_1} \left( \Fa_{\mu_2\mu_3} \bullet \left( \Delta \TB_{\mu_4\mu_5} \bullet \Fb_{\mu_6 \mu_7 \mu_8} \right) \right) \right] \epsilon^{\mu_1\ldots \mu_8} \,. \label{eq:VaryTopTerm}
 \end{split}
\end{equation}
As noted earlier, when the trombone is non-vanishing, there is no action principle because we cannot integrate by parts, mirroring the behaviour in gauged SUGRA \cite{LeDiffon:2008sh} and in ``gauged'' EFT \cite{Hohm:2014qga}.

\section{The action} \label{s:Action}
We now wish to write an EFT action, with a curved $\G$-structure. This has a similar form to seven-dimensional maximal gauged supergravity, with an ``external'' seven-dimensional metric $g_{\mu\nu}$ with vielbein $e^{\bmu}{}_{\mu}$. Under generalised diffeomorphisms, this external vielbein transforms as a scalar of weight $1/5$, i.e.
\begin{equation}
 \gL_\Lambda e^{\bmu}{}_{\mu} = \frac12 \Lambda^{ab} \nabla_{ab} e^{\bmu}{}_{\mu} + \frac{1}{10} e^{\bmu}{}_{\mu} \nabla_{ab} \Lambda^{ab} \,. \label{eq:GenLieVielbein}
\end{equation}
In addition there are 14 scalars parameterising the coset space $\frac{\SL{5}}{\SO{5}}$. We can write these in terms of the generalised metric
\begin{equation}
 \gM_{ab} = \tE^{\bba}{}_a \tE^{\bbb}{}_b \delta_{\bba\bbb} \,,
\end{equation}
where $a$ transforms under $\SL{5}$ and $\bba$ transforms under $\SO{5}$. Note that the structure group can always be reduced to its maximal compact subgroup, thus in this case from $\G$ to $\SO{5}$, so that the existence of $\gM_{ab}$ does not impose further restrictions on the 10-dimensional manifold. Finally there are also the field strengths of the tensor hierarchy, which have been described in detail in the preceding section \ref{s:TH}.

Schematically, the action takes the form
\begin{equation}
 S = S_{EH} + S_{SK} + S_{GK} + S_{top} + S_{pot} \,. \label{eq:action}
\end{equation}
Here, we have
\begin{itemize}
 \item $S_{EH}$ is an Einstein-Hilbert-like term, involving the $\D_{\mu}$ derivative, which is thus invariant under generalised diffeomorphisms,
 \item $S_{SK}$ is the kinetic term for the scalars $\gM_{ab}$,
 \item $S_{GK}$ contains the kinetic terms for the gauge fields of the tensor hierarchy,
 \item $S_{top}$ is the topological term, see \ref{s:TopTerm},
 \item $S_{pot}$ is the potential term, written completely in terms of $g_{\mu\nu}$ and $\gM_{ab}$.
\end{itemize}
Apart from the potential, the various terms appearing in the action \eqref{eq:action} are very similar to the usual EFT construction, see for example the original discussion in \cite{Hohm:2013vpa} and the specific example of $\SLf$ in \cite{Musaev:2015ces}, and so we will keep their discussion brief. Each term is manifestly invariant under generalised diffeomorphisms and $\GL{10}$-diffeomorphisms, but not under external diffeomorphisms, which act as follows
\begin{equation}
 \begin{split}
  \delta_\xi g_{\mu\nu} &= \xi^\rho \D_\rho g_{\mu\nu} + \D_\mu \xi^\rho g_{\rho\nu} + \D_\nu \xi^\rho g_{\mu\rho} \,, \\
  \delta_\xi \gM_{ab} &= \xi^\rho \D_\rho \gM_{ab} \,, \\
  \delta_\xi \Aa_\mu{}^{ab} &= \xi^\nu \Fa_{\nu\mu}{}^{ab} + \gM^{ac} \gM^{bd} g_{\mu\nu} \nabla_{cd} \xi^\nu \,, \\
  \Delta_\xi \Ab_{\mu\nu,a} &= \xi^\lambda \Fb_{\lambda\mu\nu,a} \,, \\
  \Delta_\xi \Ac_{\mu\nu\rho}{}^a &= \xi^\lambda \Fc_{\lambda\mu\nu\rho}{}^a \,. \label{eq:ExtDiffeos}
 \end{split}
\end{equation}
Here $\xi^\mu(x,Y)$ can depend on both the external and the extended coordinates. This is why we use $\D_\mu$, the covariant external derivative introduced in section \ref{s:TH}. For example $\D_\mu$ acts on $g_{\mu\nu}$ as
\begin{equation}
 \D_\mu g_{\nu\rho} = \partial_\mu g_{\nu\rho} - \gL_{\Aa_\mu} g_{\nu\rho} \,.
\end{equation}
The variations \eqref{eq:ExtDiffeos} are the $\GL{10}$-covariant generalisation of \cite{Hohm:2013vpa}.

We further take $\gE_{ab}$ to be independent of the external coordinates, $x^\mu$, so that
\begin{equation}
 \D_\mu \gE_{ab}{}^M = 0 \,.
\end{equation}
It follows that
\begin{equation}
 \left[ \D_\mu, \nabla_{ab} \right] = 0 \,.
\end{equation}
The variation of $\gE_{ab}{}^M$ also vanishes
\begin{equation}
 \delta_\xi \gE_{ab}{}^M = 0 \,.
\end{equation}

Requiring (on-shell) invariance under the external diffeomorphisms fixes the relative coefficients between the terms appearing in \eqref{eq:action}. We leave the details of the calculation to the appendix \ref{A:ExtDiffeos}.

\subsection{Covariant Einstein-Hilbert term}
Here we follow \cite{Berman:2015rcc} in constructing an Einstein-Hilbert term for the external metric $g_{\mu\nu}$ that is invariant under generalised diffeomorphisms. The alternative is to use the vielbein formalism \cite{Hohm:2013vpa}. We can define a Riemann tensor that is covariant under external diffeomorphisms, generalised diffeomorphisms and $\GL{10}$-diffeomorphisms as in the usual way, but everywhere replacing $\partial_\mu \rightarrow \D_\mu$, i.e.
\begin{equation}
 R^{\mu}{}_{\nu\rho\sigma} = \D_\rho \Gamma^\mu{}_{\nu\sigma} - \D_\sigma \Gamma^\mu{}_{\nu\rho} + \Gamma^\mu{}_{\lambda\rho} \Gamma^\lambda{}_{\nu\sigma} - \Gamma^\mu{}_{\lambda\sigma} \Gamma^\lambda{}_{\nu\rho} \,,
\end{equation}
where
\begin{equation}
 \Gamma^\mu{}_{\nu\rho} = g^{\mu\sigma} \left( \D_{(\nu} g_{\rho)\sigma} - \frac12 \D_\sigma g_{\nu\rho} \right) \,.
\end{equation}
The Einstein-Hilbert term is then
\begin{equation}
 S_{EH} = \int \IM g^{\mu\nu} R^{\rho}{}_{\mu\rho\nu} \,.
\end{equation}
For the variation under external diffeomorphisms, see appendix \ref{A:ExtDiffeos}, it useful to integrate all terms involving second-order derivatives by parts to obtain -- up to boundary terms --
\begin{equation}
 \begin{split}
  S_{EH} &= \int \IM \left[ \frac12 \D_\mu g^{\mu\nu} \D_\nu \ln g + \frac14 g^{\mu\nu} \D_\mu \ln g \, \D_\nu \ln g \right. \\
  & \quad \left. + \frac14 g^{\mu\nu} \D_\mu g^{\rho\sigma} \D_\nu g_{\rho\sigma} - \frac12 g^{\mu\nu} \D_\mu g^{\rho\sigma} \D_\rho g_{\nu\sigma} \right] \,. \label{eq:SEH}
 \end{split}
\end{equation}

\subsection{Kinetic terms}
The kinetic term for the scalar is again defined simply by replacing the usual partial derivative with a covariant derivative $\partial_\mu \rightarrow \D_\mu$, so that
\begin{equation}
 S_{SK} = \frac14 \int \IM g^{\mu\nu} \D_\mu \gM^{ab} \D_\nu \gM_{ab} \,. \label{eq:SK}
\end{equation}
The coefficient $\frac14$ is required to ensure invariance under external diffeomorphisms.

For the gauge potentials we use the field strengths defined in section \ref{s:TensorHier} but we only introduce kinetic terms for $\Fa_{\mu\nu}$ and $\Fb_{\mu\nu\rho}$ as the higher forms can be dualised to just these two. We obtain the action
\begin{equation}
 S_{GK} = -\frac18 \int \IM \left( \Fa_{\mu\nu}{}^{ab} \Fa^{\mu\nu,cd} \gM_{ac} \gM_{bd} + \frac{2}{3} \Fb_{\mu\nu\rho,a} \Fb^{\mu\nu\rho}{}_{b} \gM^{ab} \right) \,, \label{eq:GK}
\end{equation}
where the factors $-\frac18$ and $\frac23$ are required to ensure invariance under external diffeomorphisms. At this point we should also highlight that the equation of motion coming from varying $\Ac_{\mu\nu\rho}$ here and in the topological term \eqref{eq:TopTerm} gives rise to a duality relation which in the locally flat case reduces to the M-theory duality between three-form and six-form and to the IIB self-duality. It takes the form
\begin{equation}
 \epsilon^{\mu_4\ldots\mu_7} \Fc_{\mu_1\ldots\mu_7}{}^a \propto e \Fb^{\mu_1\mu_2\mu_3}{}_b \gM^{ab} \,,
\end{equation}
and is required for the action to be invariant under external diffeomorphisms, see appendix \ref{A:ExtDiffeos}.

\subsection{Scalar potential}
We next consider the scalar potential. This is expressed in terms of the scalar degrees of freedom which are encapsulated in the generalised metric $\gM_{ab}$. We will calculate the scalar potential by requiring it to be invariant under generalised diffeomorphisms as well as reducing to the right supergravity action in the locally flat case and when the section condition is solved. This implies that the potential is made of two independently invariant parts
\begin{equation}
 S_{pot} = \int \IM \left( V_1 + \frac18 \gM^{ac} \gM^{bd} \tnabla_{ab} g^{\mu\nu} \tnabla_{cd} g_{\mu\nu} \right) \,,
\end{equation}
where $V_1$ depends only on $\gM$ and $\tnabla$ is a connection under the generalised Lie derivative \eqref{eq:GenLieDerivWeight} and defined as
\begin{equation}
 \tnabla_{ab} V^c = \nabla_{ab} V^c + \tGamma_{ab,d}{}^c V^d - w \gamma_{ab} V^c \,,
\end{equation}
for a vector $V^c$ of weight $w$. We will use a Weitzenb\"ock-like connection
\begin{equation}
 \tGamma_{ab,c}{}^d = - \tE^{\bbc}{}_c \nabla_{ab} \gE_{\bbd}{}^c \,, \qquad \gamma_{ab} = \frac{5}{7} D_{ab} \ln e \,, \label{eq:FluctConnection}
\end{equation}
and derive $V_1$ in terms of the generalised torsion of this connection, mirroring the construction in \cite{Berman:2013uda,Blair:2014zba}.\footnote{This is related to the flux formulation of DFT and EFT but is different to the construction used in \cite{Bosque:2015jda}.} The generalised torsion is defined as
\begin{equation}
 \left( \gL_\Lambda^{\tnabla} - \gL_\Lambda \right) V^a = \frac12 T_{bc,d}{}^a \Lambda^{bc} V^d - \frac12 w \tT_{bc} \Lambda^{bc} V^a \,, \label{eq:Torsion}
\end{equation}
where $\tT_{ab}$ is an irrep of the generalised torsion. Explicitly, we find
\begin{equation}
 T_{ab,c}{}^d = \tGamma_{ab,c}{}^d - \frac45 \delta_c^d \tGamma_{e[a,b]}{}^e + 2 \tGamma_{c[a,b]}{}^d - 2 \tGamma_{ce,[a}{}^e \delta_{b]}^d - \frac{2}{25} \delta_c^d \gamma_{ab} - \frac25 \gamma_{c[a} \delta_{b]}^d \,.
\end{equation}

From \eqref{eq:Torsion} and \eqref{eq:FluctConnection} one can see that the torsion is invariant under $\GL{10}$ diffeomorphisms and transforms as a tensor of weight $-\frac15$ under generalised diffeomorphisms. However, for the connection \eqref{eq:FluctConnection} this torsion is not $\SO{5}$ invariant. We will return to this shortly to determine the potential uniquely. Let us first decompose $T_{ab,c}{}^d$ into its irreps
\begin{equation}
 T_{ab,c}{}^d = \frac12 \delta_{[a}^d \tS_{b]c} + \frac12 \epsilon_{abcef} \tZ^{ef,d} - \frac{1}{27} \left( 25\, \delta_{[a}^d \tT_{b]c} + 5\, \delta_c^d \tT_{ab} \right) \,.
\end{equation}
Explicitly, these are given by
\begin{equation}
 \begin{split}
  \tS_{ab} &= T_{c(a,b)}{}^c = 4 \tGamma_{c(a,b)}{}^c \,, \\
  \tZ^{ab,c} &= \frac{1}{3!} \epsilon^{abdef} T_{de,f}{}^c = \frac12 \epsilon^{abdef} \tGamma_{de,f}{}^c - \frac12 \epsilon^{abcde} \tGamma_{[fd,e]}{}^f \,, \\
  \tT_{ab} &= - \frac53 T_{c[a,b]}{}^c = \frac65 \gamma_{ab} + \tGamma_{e[a,b]}{}^e \,,
 \end{split}
\end{equation}
and live in the $\mathbf{15} \oplus \mathbf{40}' \oplus \mathbf{10}$, just as the embedding tensor does \cite{Samtleben:2005bp}. We can now construct six independent generalised diffeomorphism scalar densities
\begin{align}
 A &= \tS_{ab} \tS_{cd} \gM^{ac} \gM^{bd} \,, & B &= \left( \tS_{ab} \gM^{ab} \right)^2 \,, \nonumber\\
 C &= \gM^{ac} \gM^{bd} \tT_{ab} \tT_{cd} \,, & D &= \gM_{ab} \gM_{cd} \gM_{ef} \tZ^{ac,e} \tZ^{bd,f} \,, \\
 E &= \gM_{ab} \gM_{cd} \gM_{ef} \tZ^{ac,b} \tZ^{de,f} \,, & F &= \gM^{ac} \gM^{bd} \tnabla_{ab} \tT_{cd} \,.\nonumber
\end{align}
These terms do not individually form $\SO{5}$ scalars. However, the following combination does
\begin{equation}
 \begin{split}
  V_1 &= - \frac{1}{16} \gM^{ac} \gM^{bd} \tS_{ab} \tS_{cd} + \frac{1}{32} \gM^{ac} \gM^{bd} \tS_{ac} \tS_{bd} - \frac{5}{12} \gM^{ac} \gM^{bd} \tT_{ab} \tT_{cd} \\
  & \quad - \frac12 \gM_{ab} \gM_{cd} \gM_{ef} \tZ^{ac,e} \tZ^{bd,f} + \frac12 \gM_{ab} \gM_{cd} \gM_{ef} \tZ^{ac,b} \tZ^{ed,f} - \gM^{ac} \gM^{bd} \tnabla_{ab} \tT_{cd} \,. \label{eq:V1EmbTensor}
 \end{split}
\end{equation}
One can see this by explicitly rewriting this in terms of the generalised metric. The details of this calculation can be found in appendix \ref{A:ScalarPot}. When the background trombone vanishes, $\tau_{ab} = 0$, then we can write \eqref{eq:V1EmbTensor} up to the section condition \eqref{eq:SecCon}, the quadratic and linear constraints \eqref{eq:QC}, \eqref{eq:LCTau} and the relationship \eqref{eq:AnholCoeff} as follows
\begin{dmath}
 S_{pot} = \int \IM \Bigg( \frac18 \gM^{ac} \gM^{bd} \nabla_{ab} g^{\mu\nu} \nabla_{cd} g_{\mu\nu} - \frac{5}{14} \gM^{ac} \gM^{bd} D_{ab} \ln e \, D_{cd} \ln e - \frac{12}{7} \gM^{ac} D_{ab} \gM^{bd} D_{cd} \ln e - \frac67 \gM^{ac} \gM^{bd} \nabla_{ab} D_{cd} \ln e - \frac12 \nabla_{ab} \gM^{ac} \nabla_{cd} \gM^{bd} + \frac18 \gM^{ac} \gM^{bd} \nabla_{ab} \gM^{ef} \nabla_{cd} \gM_{ef} + \frac12 \gM^{ac} \gM^{bd} \nabla_{ab} \gM^{ef} \nabla_{ec} \gM_{df} - \gM^{ac} \nabla_{ab} \nabla_{cd} \gM^{bd} + \gM^{ac} \gM^{bd} \Big( - \frac12 \Conn_{ae,c}{}^e \Conn_{bf,d}{}^f - \frac12 \Conn_{ae,c}{}^f \Conn_{bf,d}{}^e + \frac12 \Conn_{ae,b}{}^e \Conn_{df,c}{}^f - \frac12 \Conn_{ae,b}{}^f \Conn_{df,c}{}^e - \Conn_{ae,d}{}^f \Conn_{fc,b}{}^e - \Conn_{ae,f}{}^e \Conn_{cd,b}{}^f \Big) \Bigg) \,. \label{eq:MetricPotential}
\end{dmath}

This form of the scalar potential is manifestly $\SO{5}$ invariant but no longer manifestly invariant under generalised diffeomorphisms. The appearance of the connection components $\Conn_{ab,c}{}^d$ is similar to the structure of the scalar potential of $\DWZW$ \cite{Blumenhagen:2014gva,Blumenhagen:2015zma}. It would also be interesting to see this potential derived using a torsion-free connection, for example following \cite{Cederwall:2013naa,Hohm:2013vpa,Park:2013gaj} for the fluctuations.

Before moving on, let us give the action in a form where the boundary terms have been integrated by parts (in the case when $\tau_{ab} = 0$ vanishes)
\begin{dmath}
 S_{pot} = \int \IM \Bigg( \frac18 \gM^{ac} \gM^{bd} \nabla_{ab} g^{\mu\nu} \nabla_{cd} g_{\mu\nu} + \frac{1}{8} \gM^{ac} \gM^{bd} D_{ab} \ln g \, D_{cd} \ln g + \frac12 \gM^{ac} \nabla_{ab} \gM^{bd} \nabla_{cd} \ln g + \frac12 \nabla_{ab} \gM^{ac} \nabla_{cd} \gM^{bd} + \frac18 \gM^{ac} \gM^{bd} \nabla_{ab} \gM^{ef} \nabla_{cd} \gM_{ef} + \frac12 \gM^{ac} \gM^{bd} \nabla_{ab} \gM^{ef} \nabla_{ec} \gM_{df} + \gM^{ac} \gM^{bd} \bigg( - \frac12 \Conn_{ae,c}{}^e \Conn_{bf,d}{}^f - \frac12 \Conn_{ae,c}{}^f \Conn_{bf,d}{}^e + \frac12 \Conn_{ae,b}{}^e \Conn_{df,c}{}^f - \frac12 \Conn_{ae,b}{}^f \Conn_{df,c}{}^e - \Conn_{ae,d}{}^f \Conn_{fc,b}{}^e - \Conn_{ae,f}{}^e \Conn_{cd,b}{}^f \bigg) \Bigg) \,. \label{eq:SPot1Order}
\end{dmath}
We can also rewrite this form of the scalar potential in terms of the ``big generalised metric'' $M^{ab,cd} = 2 \gM^{a[c} \gM^{d]b}$. This allows one to compare the scalar potential to that found in the flat $\SLf$ EFT \cite{Musaev:2015ces}. The result is
\begin{equation}
 \begin{split}
  S_{pot} &= \int d^{10}Y d^7x \, \gE e \left( \frac14 M^{AB} \nabla_{A} g^{\mu\nu} \nabla_{B} g_{\mu\nu} + \frac14 M^{AB} D_{A} \ln g \, D_{B} \ln g \right. \\
  & \left. \quad + \frac12 \nabla_A M^{AB} \nabla_B \ln g + \frac{1}{12} M^{AB} \nabla_A M^{CD} \nabla_B M_{CD} - \frac{1}{2} M^{AB} \nabla_A M^{CD} \nabla_{C} M_{BD} \right) \\
  & \quad - M^{ab,cd} \left( \frac12 \omega_{ae,c}{}^e \omega_{bf,d}{}^f + \frac12 \omega_{ae,f}{}^e \omega_{cd,b}{}^f + \frac12 \omega_{ae,c}{}^f \omega_{bf,d}{}^e - \omega_{ae,d}{}^f \omega_{fc,b}{}^e \right) \,. \label{eq:BigAction}
 \end{split}
\end{equation}
Here we label antisymmetric pairs of indices as $A = [ab]$, $A = 1, \ldots, 10$ and every contraction of these 10-d indices $A, B$ comes with a factor of $\frac12$ when written in terms of the $\SLf$ indices. For example, the first term would read
\begin{equation}
 \frac14 M^{AB} \nabla_{A} g^{\mu\nu} \nabla_{B} g_{\mu\nu} = \frac{1}{16} M^{ab,cd} \nabla_{ab} g^{\mu\nu} \nabla_{cd} g_{\mu\nu} \,.
\end{equation}
The details of this calculation can be found in appendix \ref{A:BigMetric}.

The first two lines in \eqref{eq:BigAction} reduce immediately to the $\SLf$ EFT action of \cite{Musaev:2015ces} when the covariant derivatives are replaced by partial derivatives, while the final line represents a non-minimal modification in the case of fluxes, just as in $\DWZW$ \cite{Blumenhagen:2014gva,Blumenhagen:2015zma,Bosque:2015jda}. It is thus easy to see that when the fluxes vanish we reproduce the usual $\SLf$ action \cite{Musaev:2015ces}. When they do not vanish but the $\G$-structure is locally flat, we expect to obtain the gauged $\SLf$ EFT action.

\section{Discussion} \label{s:Discussion}
In this paper we have shown that it is possible to define a full $\SLf$ EFT, on any 10-manifold with $\G$-structure. When the $\G$-structure is locally flat the formulation here reduces to the usual EFT formulation. Furthermore the action given here reduces exactly to the one found in the usual EFT formulation \cite{Musaev:2015ces}. The benefit of the approach here is that since we are patching the EFT not just with generalised diffeomorphisms, but with ordinary $\GL{10}$-diffeomorphisms, we can also describe non-geometric backgrounds, as first discussed in \cite{Cederwall:2014opa}. In the set-up presented here usual geometric backgrounds should then be related to 10-manifolds whose structure group can be further reduced from $\G$ to $G_{geom}$, its geometric subgroup. When this is not possible, the background would be non-geometric.

One may also wonder what the physical significance of the curvature of the $\G$-structure is. At this point we can only speculate that this may allow us to describe non-Abelian T-dualities \cite{delaOssa:1992vci}. Also, as argued in \cite{Hohm:2015ugy}, when the $\G$-structure is not locally flat, the theory is not background dependent. This is due to the fact that fluctuations about the background are described by elements of $\G$ but the background, encapsulated in the $\GL{10}$ vielbeine $\gE_{ab}{}^M$, is not.

This may suggest that we should limit ourselves to 10-manifolds with locally flat $\G$-structure. However, as discussed in \cite{Lee:2015xga} for the $\ODD$ case, this is a very restrictive requirement for the extended space. One may wonder what three- or four-dimensional manifolds can be described as solutions of the section condition on this restricted set of 10-manifolds.

An interesting contrast to the generalised geometry picture then emerges: as noted in \cite{Lee:2015xga} while for DFT / EFT, there would be restrictions on the allowed extended manifold, it is always possible to define a generalised geometry with flat $\ODD$- or $\Edd$-structure on the generalised tangent bundle of \emph{any} manifold. This may be reconcilable since the local flatness restriction applies to the full doubled space not to the physical manifold obtained after applying the section condition. Indeed, it has been observed in \cite{Hohm:2015ugy} that the perturbations around WZW-backgrounds can be adequately described by DFT. This seems to suggest that these manifolds admit a locally flat $\ODD$ structure and so would be an explicit example of backgrounds with interesting topology whose doubled space may admit locally flat $\ODD$-structure. We leave these questions open for future publications.

\section{Acknowledgements}
We have benefited from useful discussions with David Berman, Chris Blair, Martin Cederwall, Michael Haack, Jonathan Heckman, Daniel Thompson and Daniel Waldram. FH would like to thank the theory groups at Columbia University and the ITS at the CUNY graduate center for hospitality during the completion of this work. EM would like to thank the organizers of the APCTP workshop ``Duality and Novel Geometry in M-theory'' for their hospitality, while part of this work was being completed. The work of PB, EM and DL is supported by the ERC Advanced Grant ``Strings and Gravity" (Grant No. 32004). The work of FH is supported by NSF CAREER grant PHY-1452037. FH also acknowledges the support from the Bahson Fund at UNC Chapel Hill as well as the R. J. Reynolds Industries, Inc. Junior Faculty Development Award from the Office of the Executive Vice Chancellor and Provost at UNC Chapel Hill.

\appendix
\section{Closure constraints}\label{A:Closure}
We begin with \eqref{eq:GenLieDeriv}, where $\tau_{bc,d}{}^a$ is for now an unspecified constant, i.e.
\begin{equation}
 \gL_\Lambda V^{a} = \frac{1}{2} \Lambda^{bc} D_{bc} V^a + \frac{1}{5} V^a D_{bc} \Lambda^{bc} - V^b D_{bc} \Lambda^{ac} + \frac{1}{2} \tau_{bc,d}{}^a \Lambda^{bc} V^d \,, \label{eq:GenLieDerivA}
\end{equation}
where $\Lambda$ has weight $\frac15$. The algebra closes if
\begin{equation}
 \left[ \gL_{\Lambda_1} ,\, \gL_{\Lambda_2} \right] V^a = \gL_{\left[\Lambda_1, \, \Lambda_2\right]_E} V^a \,,
\end{equation}
with
\begin{equation}
 \left[ \Lambda_1 , \, \Lambda_2 \right]_E = \frac12 \left( \gL_{\Lambda_1} \Lambda_2 - \gL_{\Lambda_2} \Lambda_1 \right) \,.
\end{equation}
To keep this tractable we will call
\begin{equation}
 A^a = \left[ \gL_{\Lambda_1} ,\, \gL_{\Lambda_2} \right] V^a \,, \qquad B^a = \gL_{\left[\Lambda_1, \, \Lambda_2\right]_E} V^a \,, \qquad C^a = A^a - B^a \,.
\end{equation}
Each of these expressions will involve terms which are independent of $\tau_{ab,c}{}^d$, linear in $\tau_{ab,c}{}^d$ and quadratic in $\tau_{ab,c}{}^d$. We will denote these with the subscripts $0$, $1$ and $2$. We find for the $\tau_{ab,c}{}^d$-independent terms
\begin{equation}
 \begin{split}
  A^a_0 &= \frac18 \Lambda_1^{bc} \Lambda_2^{de} \left[ D_{bc}, D_{de} \right] V^a + \frac{1}{10} V^a \Lambda_1^{bc} D_{bc} D_{de} \Lambda_2^{de} - \frac12 V^b \Lambda_1^{de} D_{de} D_{bc} \Lambda_2^{ac} \\
  & \quad + \frac14 \Lambda_1^{de} D_{de} \Lambda_2^{bc} D_{bc} V^a - V^b D_{be} \Lambda_1^{de} D_{dc} \Lambda_2^{ac} - (1 \leftrightarrow 2) \,, \\
  B^a_0 &= \frac14 \Lambda_1^{de} D_{de} \Lambda_2^{bc} D_{bc} V^a - \frac34 \Lambda_1^{de} D_{[de} \Lambda_2^{bc} D_{bc]} V^a - \frac15 V^a \Lambda_1^{cd} D_{bc} D_{de} \Lambda_2^{be} \\
  & \quad + \frac34 V^b \Lambda_1^{de} D_{b[c} D_{de]} \Lambda_2^{ac} - \frac12 V^b \Lambda_1^{de} D_{bc} D_{de} \Lambda_2^{ac} + \frac34 V^b \Lambda_1^{ac} D_{b[c} D_{de]} \Lambda_2^{de} \\
  & \quad - \frac32 V^b D_{[bc} \Lambda_1^{de} D_{de]} \Lambda_2^{ac} + \frac12 V^b D_{be} \Lambda_1^{de} D_{cd} \Lambda_2^{ac} - \left( 1 \leftrightarrow 2 \right) \,, \\
 \end{split}
\end{equation}
To simplify these expressions, note that
\begin{equation}
 D_{[ab} D_{cd]} = \frac12 D_{a[b} D_{cd]} - \frac12 D_{[cd} D_{b]a} = D_{a[b} D_{cd]} - \frac12 \left[ D_{a[b}, D_{cd]} \right] \,.
\end{equation}
Hence we can write
\begin{equation}
 D_{a[b} D_{cd]} = D_{[ab} D_{cd]} + \frac12 \left[D_{a[b}, D_{cd]} \right] \,, \qquad D_{[cd} D_{|a|b]} = D_{[ab} D_{cd]} - \frac12 \left[ D_{a[b}, D_{cd]} \right] \,. \label{eq:34Antisymm}
\end{equation}
Using these identities we find that the difference $C^a$ is given by
\begin{equation}
 \begin{split}
  C^a_0 &= \frac18 \Lambda_1^{cd} \Lambda_2^{ef} \left[ D_{cd}, D_{ef} \right] V^a - \frac{1}{10} V^a \Lambda_1^{de} \left[ D_{bc}, D_{de} \right] \Lambda_2^{bc} + \frac{3}{10} V^a \Lambda_1^{de} D_{[de} D_{bc]} \Lambda_2^{bc} \\
  & \quad - \frac38 V^b \Lambda_1^{de} \left[ D_{de}, D_{bc} \right] \Lambda_2^{ac} - \frac14 V^b \Lambda_1^{de} \left[ D_{be}, D_{cd} \right] \Lambda_2^{ac} - \frac34 V^b \Lambda_1^{de} D_{[bc} D_{de]} \Lambda_2^{ac} \\
  & \quad + \frac14 V^b \Lambda_1^{ac} \left[ D_{bd}, D_{ce} \right] \Lambda_2^{de} - \frac18 V^b \Lambda_1^{ac} \left[ D_{bc}, D_{de} \right] \Lambda_2^{de} - \frac34 V^b \Lambda_1^{ac} D_{[bc} D_{de]} \Lambda_2^{de} \\
  & \quad + \frac34 \Lambda_1^{de} D_{[de} \Lambda_2^{bc} D_{bc]} V^a - \frac32 V^b D_{[bc} \Lambda_1^{de} D_{de]} \Lambda_2^{ac} - \left( 1 \leftrightarrow 2 \right) \,.
 \end{split}
\end{equation}

Let us now turn to the terms linear in $\tau_{ab,c}{}^d$, given by
\begin{equation}
 \begin{split}
  A^a_1 &= \frac14 V^f \Lambda_1^{de} D_{de} \Lambda_2^{bc} \tau_{bc,f}{}^a - \frac12 V^f \Lambda_1^{de} D_{fc} \Lambda_2^{bc} \tau_{de,b}{}^a + \frac12 V^f \Lambda_1^{bc} D_{de} \Lambda_2^{ae} \tau_{bc,f}{}^d - \left( 1 \leftrightarrow 2 \right) \,, \\
  B^a_1 &= \frac{1}{40} \tau_{de} \Lambda_1^{de} \Lambda_2^{bc} D_{bc} V^a - \frac{1}{40} \tau_{de} \Lambda_1^{bc} \Lambda_2^{de} D_{bc} V^a + \frac18 \tau_{de,[f}{}^{[b} \delta_{g]}{}^{c]} \Lambda_1^{de} \Lambda_2^{fg} D_{bc} V^a - \frac18 \tau_{de,[f}{}^{[b} \delta_{g]}{}^{c]} \Lambda_1^{fg} \Lambda_2^{de} D_{bc} V^a \\
  & \quad - \frac{1}{50} V^a \tau_{de} \Lambda_1^{bc} D_{bc} \Lambda_2^{de} + \frac{1}{50} V^a \tau_{de} \Lambda_1^{de} D_{bc} \Lambda_2^{bc} + \frac{1}{10} V^a \Lambda_1^{de} \tau_{de,f}{}^c D_{bc} \Lambda_2^{bf} - \frac{1}{10} V^a \Lambda_1^{bf} \tau_{de,f}{}^c D_{bc} \Lambda_2^{de} \\
  & \quad - \frac14 V^b \tau_{de,f}{}^c \Lambda_1^{de} D_{bc} \Lambda_2^{af} - \frac{1}{10} \tau_{de} V^b \Lambda_1^{de} D_{bc} \Lambda_2^{ac} + \frac14 V^b \tau_{de,f}{}^c \Lambda_1^{af} D_{bc} \Lambda_2^{de} + \frac{1}{10} V^b \tau_{de} \Lambda_1^{ac} D_{bc} \Lambda_2^{de} \\
  & \quad + \tau_{de,f}{}^a \left( \frac14 V^b \Lambda_1^{de} D_{bc} \Lambda_2^{cf} - \frac14 V^b \Lambda_1^{cf} D_{bc} \Lambda_2^{de} - \frac18 V^f \Lambda_1^{de} D_{bc} \Lambda_2^{bc} + \frac18 V^f \Lambda_1^{bc} D_{bc} \Lambda_2^{de} \right. \\
  & \quad \left. + \frac12 V^f \Lambda_1^{be} D_{bc} \Lambda_2^{dc} \right) - \left( 1 \leftrightarrow 2 \right) \,, \\
  C^a_1 &= \frac18 \left( - \frac{1}{5} \tau_{de} \Lambda_1^{de} \Lambda_2^{bc} + \frac{1}{5} \tau_{de} \Lambda_1^{bc} \Lambda_2^{de} - \tau_{de,[f}{}^{[b} \delta_{g]}{}^{c]} \Lambda_1^{de} \Lambda_2^{fg} + \tau_{de,[f}{}^{[b} \delta_{g]}{}^{c]} \Lambda_1^{fg} \Lambda_2^{de} \right) D_{bc} V^a \\
  & \quad + \frac{1}{10} V^a \left( \frac{1}{5} \tau_{de} \Lambda_1^{bc} D_{bc} \Lambda_2^{de} - \frac{1}{5} \tau_{de} \Lambda_1^{de} D_{bc} \Lambda_2^{bc} - \Lambda_1^{de} \tau_{de,f}{}^c D_{bc} \Lambda_2^{bf} + \Lambda_1^{bf} \tau_{de,f}{}^c D_{bc} \Lambda_2^{de} \right) \\
  & \quad + V^b \Lambda_1^{de} \left( \frac14 \tau_{de,f}{}^c D_{bc} \Lambda_2^{af} + \frac12 \tau_{de,b}{}^f D_{fc} \Lambda_2^{ac} + \frac{1}{10} \tau_{de} D_{bc} \Lambda_2^{ac} \right) \\
  & \quad + V^b D_{bc} \Lambda_2^{de} \left( - \frac14 \tau_{de,f}{}^c \Lambda_1^{af} - \frac{1}{10} \tau_{de} \Lambda_1^{ac} \right) + \tau_{de,f}{}^a \left( \frac14 V^b \Lambda_1^{de} D_{bc} \Lambda_2^{cf} + \frac14 V^b \Lambda_1^{cf} D_{bc} \Lambda_2^{de} \right. \\
  & \quad \left. + \frac18 V^f \Lambda_1^{de} D_{bc} \Lambda_2^{bc} + \frac18 V^f \Lambda_1^{bc} D_{bc} \Lambda_2^{de} + \frac12 V^f \Lambda_1^{be} D_{bc} \Lambda_2^{cd} \right) - \left( 1 \leftrightarrow 2 \right) \,.
 \end{split}
\end{equation}
For later on, it will be convenient to further split $C_1^a = D_1^a + E_1^a + F_1^a$ with
\begin{equation}
 \begin{split}
  D^a_1 &= \tau_{de,f}{}^a \left( \frac14 V^b \Lambda_1^{de} D_{bc} \Lambda_2^{cf} + \frac14 V^b \Lambda_1^{cf} D_{bc} \Lambda_2^{de} + \frac18 V^f \Lambda_1^{de} D_{bc} \Lambda_2^{bc} + \frac18 V^f \Lambda_1^{bc} D_{bc} \Lambda_2^{de} \right. \\
  & \left. \quad + \frac12 V^f \Lambda_1^{be} D_{bc} \Lambda_2^{cd} \right) - \left( 1 \leftrightarrow 2 \right) \\
  E_1^a &= V^b \Lambda_1^{de} \left( \frac14 \tau_{de,f}{}^c D_{bc} \Lambda_2^{af} + \frac12 \tau_{de,b}{}^f D_{fc} \Lambda_2^{ac} + \frac{1}{10} \tau_{de} D_{bc} \Lambda_2^{ac} \right) \\
  & \quad + V^b D_{bc} \Lambda_2^{de} \left( - \frac14 \tau_{de,f}{}^c \Lambda_1^{af} - \frac{1}{10} \tau_{de} \Lambda_1^{ac} \right) - \left( 1 \leftrightarrow 2 \right) \\
  F_1^a &= \frac18 \left( - \frac{1}{5} \tau_{de} \Lambda_1^{de} \Lambda_2^{bc} + \frac{1}{5} \tau_{de} \Lambda_1^{bc} \Lambda_2^{de} - \tau_{de,[f}{}^{[b} \delta_{g]}{}^{c]} + \tau_{de,[f}{}^{[b} \delta_{g]}{}^{c]} \Lambda_1^{fg} \Lambda_2^{de} \right) D_{bc} V^a \\
  & \quad + \frac{1}{10} V^a \left( \frac{1}{5} \tau_{de} \Lambda_1^{bc} D_{bc} \Lambda_2^{de} - \frac{1}{5} \tau_{de} \Lambda_1^{de} D_{bc} \Lambda_2^{bc} - \Lambda_1^{de} \tau_{de,f}{}^c D_{bc} \Lambda_2^{bf} + \Lambda_1^{bf} \tau_{de,f}{}^c D_{bc} \Lambda_2^{de} \right) \\
  &\quad - \left( 1 \leftrightarrow 2 \right) \,.
 \end{split}
\end{equation}

Finally, the terms quadratic in $\tau_{ab,c}{}^d$ are given by
\begin{equation}
 \begin{split}
  A^a_2 &= \frac18 V^f \Lambda_1^{bc} \Lambda_2^{de} \left( \tau_{bc,g}{}^a \tau_{de,f}{}^g - \tau_{de,g}{}^a \tau_{bc,f}{}^g \right) - \left( 1 \leftrightarrow 2 \right) \,, \\
  B^a_2 &= \frac18 V^f \Lambda_1^{bc} \Lambda_2^{de} \left( \tau_{dg,f}{}^a \tau_{bc,e}{}^g - \tau_{bg,f}{}^a \tau_{de,c}{}^g + \frac15 \tau_{de,f}{}^a \tau_{bc} - \frac15 \tau_{bc,f}{}^a \tau_{de} \right) - \left( 1 \leftrightarrow 2 \right) \,.
 \end{split}
\end{equation}
Let us begin with the quadratic terms. Their difference is given by
\begin{equation}
 \begin{split}
  C_2^a &= \frac18 V^f \Lambda_1^{bc} \Lambda_2^{de} \bigg( \tau_{bc,g}{}^a \tau_{de,f}{}^g - \tau_{de,g}{}^a \tau_{bc,f}{}^g - \tau_{dg,f}{}^a \tau_{bc,e}{}^g + \tau_{bg,f}{}^a \tau_{de,c}{}^g \bigg. \\
  & \quad \bigg. - \frac15 \tau_{de,f}{}^a \tau_{bc} + \frac15 \tau_{bc,f}{}^a \tau_{de} \bigg) - \left( 1 \leftrightarrow 2 \right) \\
  &= \frac18 V^f \Lambda_1^{bc} \Lambda_2^{de} \bigg( \hat{\tau}_{bc,g}{}^a \hat{\tau}_{de,f}{}^g - \hat{\tau}_{de,g}{}^a \hat{\tau}_{bc,f}{}^g - \hat{\tau}_{dg,f}{}^a \hat{\tau}_{bc,e}{}^g + \hat{\tau}_{bg,f}{}^a \hat{\tau}_{de,c}{}^g \bigg. \\
  & \quad \bigg. + \frac15 \delta^a_f \hat{\tau}_{bc,e}{}^g \tau_{dg} - \frac15 \delta^a_f \hat{\tau}_{de,c}{}^g \tau_{bg} \bigg) - \left( 1 \leftrightarrow 2 \right) \,,
 \end{split}
\end{equation}
where $\hat{\tau}_{ab,c}{}^d = \tau_{ab,c}{}^d + \frac15 \delta^d_c \tau_{ab}$. The first line is proportional to the quadratic constraint of gauged supergravities and thus vanishes if we impose that quadratic constraint \eqref{eq:QC}
\begin{equation}
 2 \hat{\tau}_{ab,[c}{}^h \hat{\tau}_{d]h,e}{}^f - \hat{\tau}_{ab,e}{}^h \hat{\tau}_{cd,h}{}^f + \hat{\tau}_{ab,h}{}^f \hat{\tau}_{cd,e}{}^h = 0 \,. \label{eq:AQC}
\end{equation}
The second line can also be shown to vanish using the quadratic constraint, if we also impose the linear constraint \eqref{eq:QCLin}. To see this, note that the second line can be written as
\begin{equation}
 \begin{split}
  \Delta &= \frac1{40} V^a \Lambda_1^{bc} \Lambda_2^{de} \left( \hat{\tau}_{bc,e}{}^g \tau_{dg} - \hat{\tau}_{de,c}{}^g \tau_{bg} \right) - \left( 1 \leftrightarrow 2 \right) \\
  &= \frac1{40} V^a \Lambda_1^{bc} \Lambda_2^{de} \Delta_{bc,de} \,,
 \end{split}
\end{equation}
with $\Delta_{ab,cd} \in \mathbf{45}$. Thus we can also write it as
\begin{equation}
 \begin{split}
  \Delta_{a}{}^{bc} &= \frac1{3!} \epsilon^{bcdef} \Delta_{ad,ef} \,, \\
  &= \epsilon^{bcdef} \left( \frac13 \tau_{eg} \hat{\tau}_{ad,f}{}^g - \frac16 \tau_{ag} \hat{\tau}_{ef,d}{}^g + \frac16 \tau_{dg} \hat{\tau}_{ef,a}{}^g \right) \\
  &= 0 \,,
 \end{split}
\end{equation}
which vanishes by the quadratic constraint \eqref{eq:AQC}. Thus we have $C^a_2 = 0$.

Let us now look at the $D_{bc} V^a$ terms in $C^a_0$ and $C^a_1$. For these to vanish we find
\begin{equation}
 \begin{split}
  0 &= \Lambda_1^{bc} \Lambda_2^{de} \left( \left[ D_{bc}, D_{de} \right] V^a - \left( \tau_{bc,[d}{}^{[f} \delta_{e]}{}^{g]} + \tau_{de,[b}{}^{[f} \delta_{c]}{}^{g]} + \frac15 \tau_{bc} \delta_{de}^{fg} - \frac15 \tau_{de} \delta_{bc}^{fg} \right) D_{fg} V^a \right) \,.
 \end{split}
\end{equation}
Thus we are lead to impose
\begin{equation}
 \left[ D_{bc}, D_{de} \right] = \left( \tau_{bc,[d}{}^{[f} \delta_{e]}{}^{g]} - \tau_{de,[b}{}^{[f} \delta_{c]}{}^{g]} + \frac15 \tau_{bc} \delta_{de}^{fg} - \frac15 \tau_{de} \delta_{bc}^{fg} \right) D_{fg} \,, \label{eq:ACommRel}
\end{equation}
This also ensures that the terms involving $V^a$ without derivatives cancel up to the term
\begin{equation}
 \frac{3}{10}  V^a \Lambda_1^{de} D_{[de} D_{bc]} \Lambda_2^{bc} \,,
\end{equation}
of $C^a_0$, which will have to be cancelled by the remaining terms. The remaining cancellations are ensured by the linear constraint, which implies that only the $\mathbf{15}$, $\mathbf{40}'$ and $\mathbf{10} \subset \mathbf{10} \otimes \mathbf{24}$ are non-zero. Thus, we can write
\begin{equation}
 \tau_{ab,c}{}^d = \frac12 \delta_{[a}^d S_{b]c} + \frac12 \epsilon_{abcef} Z^{ef,d} + \frac{2}{15} \delta_c^d \tau_{ab} + \frac23 \delta^d_{[a} \tau_{b]c}
\end{equation}

Let us now show that with these representations $C^a$ vanishes, up to the constraint \eqref{eq:IntertwineConstraint} and the section condition \eqref{eq:SecCon}. We will do so by considering the individual representations in turn. Let us begin with the $\mathbf{15}$, so that we will for now set
\begin{equation}
 \tau_{ab,c}{}^d\vert_{15} = \frac12 \delta_{[a}^d S_{b]c} \,.
\end{equation}
Then we find for $C_1^a$
\begin{equation}
 \begin{split}
  C_1^a\vert_{15} &= W^b \left( \frac18 \Lambda_1^{ef} S_{cf} D_{be} V^{ac} - \frac14 \Lambda_1^{ef} S_{bf} S_{cf} D_{ce} \Lambda_2^{ac} - \frac18 \Lambda_1^{ac} S_{cf} D_{be} \Lambda_2^{ef} \right) \\
  & \quad + W^b \Lambda_1^{ef} \left( \frac18 S_{cf} D_{be} \Lambda_2^{ac} + \frac{1}{16} S_{cb} D_{ef} \Lambda_2^{ac} - \frac18 S_{bf} D_{ec} \Lambda_2^{ac} \right) \\
  & \quad + W^b \Lambda_1^{ac} \left( \frac18 S_{cf} D_{be} \Lambda_2^{ef} + \frac{1}{16} S_{bc} D_{ef} \Lambda_2^{ef} + \frac18 S_{bf} D_{ce} \Lambda_2^{ef} \right) - \left( 1 \leftrightarrow 2 \right) \\
  &= W^b \Lambda_1^{ef} \left( \frac14 S_{cf} D_{be} \Lambda_2^{ac} + \frac{1}{16} S_{bc} D_{ef} \Lambda_2^{ac} - \frac18 S_{bf} D_{ce} \Lambda_2^{ac} \right) \\
  & \quad + W^b \Lambda_1^{ac} \left( \frac{1}{16} S_{bc} D_{ef} \Lambda_2^{ef} + \frac18 S_{bf} D_{ce} \Lambda_2^{ef} \right) - \left( 1 \leftrightarrow 2 \right) \,.
 \end{split}
\end{equation}
In the first equality, the first line comes from the terms of $C_1^a$ where the free index $a$ is on $\Lambda_1 D \Lambda_2^{ac}$ and $\Lambda_1^{ac} D \Lambda_2$, while the second and third line come from $D_1^a\vert_{15}$.
The totally antisymmetric terms such as $D_{[ab} \otimes D_{cd]}$ cannot contribute to the $\mathbf{15}$ so we only need to consider the commutator terms of $C_0^a$. They are given by
\begin{equation}
 \begin{split}
  C_0^a\vert_{15} &= V^b \Lambda_1^{ef} \left( - \frac{3}{16} D_{be} \Lambda_2^{ac} + \frac{3}{16} S_{bd} D_{ce} \Lambda_2^{ac} - \frac{1}{16} S_{ef} D_{be} \Lambda_2^{ac} - \frac{1}{16} S_{bc} D_{ef} \Lambda_2^{ac} - \frac{1}{16} S_{bf} D_{ce} \Lambda_1^{ac} \right) \\
  & \quad + V^b \Lambda_1^{ac} \left( \frac18 S_{cf} D_{be} \Lambda_2^{ef} - \frac{1}{16} S_{bf} D_{ce} \Lambda_2^{ef} + \frac{1}{16} S_{cf} D_{be} \Lambda_2^{ef} - \frac{1}{16} S_{bc} D_{ef} \Lambda_2^{ef} - \frac{1}{16} S_{bf} D_{ce} \Lambda_2^{ef} \right) \\
  & \quad - \left( 1 \leftrightarrow 2 \right) \\
  &= V^b \Lambda_1^{ef} \left( - \frac14 S_{cf} D_{be} \Lambda_2^{ac} + \frac18 S_{bf} D_{ce} \Lambda_2^{ac} - \frac{1}{16} S_{bc} D_{ef} \Lambda_2^{ac} \right) \\
  & \quad V^b \Lambda_1^{ac} \left( - \frac{1}{16} S_{bc} D_{ef} \Lambda_2^{ef} - \frac18 S_{bf} D_{be} \Lambda_2^{ef} \right) - \left( 1 \leftrightarrow 2 \right) \,.
 \end{split}
\end{equation}
We see that $C^a\vert_{15} = 0$ as required.

Let us now turn to the $\mathbf{40}'$. For that write
\begin{equation}
 \tau_{ab,c}{}^d\vert_{40'} = \frac12 \epsilon_{abcef} Z^{ef,d} \,.
\end{equation}
We start with $D_1^a\vert_{40'}$. Note that
\begin{equation}
 \begin{split}
  5! A_{def} V^{[b} \Lambda_1^{de} D_{bc} \Lambda_2^{cf]} &= 12 A_{def} \left( 2 V^b \Lambda_1^{de} D_{bc} \Lambda_2^{cf} + 2 V^b \Lambda_1^{cf} D_{bc} \Lambda_2^{de} \right. \\
  & \quad \left. + 4 V^f \Lambda_1^{be} D_{bc} \Lambda_2^{cd} + V^f \Lambda_1^{de} D_{bc} \Lambda_2^{bc} + V^f \Lambda_1^{bc} D_{bc} \Lambda_2^{de} \right) \\
  &= A_{def} \epsilon^{bdecf} \epsilon_{ghijk} V^g \Lambda_1^{hi} D_{bc} \Lambda_2^{jk} \,.
 \end{split}
\end{equation}
where $A_{def}$ is totally antisymmetric in its indices. This is exactly the form of the terms in $D_a^1\vert_{40'}$ so we see that
\begin{equation}
 \begin{split}
  D_1^a\vert_{40'} &= \tau_{de,f}{}^a\vert_{40'} \left( \frac14 V^b \Lambda_1^{de} D_{bc} \Lambda_2^{cf} + \frac14 V^b \Lambda_1^{cf} D_{bc} \Lambda_2^{de} + \frac18 V^f \Lambda_1^{de} D_{bc} \Lambda_2^{bc} \right. \\
  & \quad \left. + \frac18 V^f \Lambda_1^{bc} D_{bc} \Lambda_2^{de} + \frac12 V^f \Lambda_1^{be} D_{bc} \Lambda_2^{cd} \right) - \left( 1 \leftrightarrow 2 \right) \\
  &= \frac12 \frac{1}{96} \epsilon_{defgh} Z^{gh,a} \epsilon^{bdecf} \epsilon_{ijklm} V^i \Lambda_1^{jk} D_{bc} \Lambda_2^{lm} - \left( 1 \leftrightarrow 2 \right) \\
  &= \frac{1}{16} Z^{bc,a} \epsilon_{defgh} V^d \Lambda_1^{ef} D_{bc} \Lambda_2^{gh} - \left( 1 \leftrightarrow 2 \right) \,, \label{eq:C1401}
 \end{split}
\end{equation}
The remaining terms of $C_1^a\vert_{40'}$ are in $E_1^a\vert_{40'}$ and are given by
\begin{equation}
 \begin{split}
  E_1^a\vert_{40'} &= V^b \Lambda_1^{de} \left( \frac14 \tau_{de,c}{}^f\vert_{40'} D_{bf} \Lambda_2^{ac} + \frac12 \tau_{de,b}{}^f\vert_{40'} D_{fc} \Lambda_2^{ac} \right) - \frac14 V^b \Lambda_1^{ac} \tau_{de,c}{}^f\vert_{40'} D_{bf} \Lambda_2^{de} \\
  & \quad - \left( 1 \leftrightarrow 2 \right) \\
  &= V^b \Lambda_1^{de} Z^{gh,f} \left( \frac18 \epsilon_{decgh} D_{bf} \Lambda_2^{ac} + \frac14 \epsilon_{debgh} D_{fc} \Lambda_2^{ac} \right) - \frac18 V^b \Lambda_1^{ac} Z^{gh,f} \epsilon_{decgh} D_{bf} \Lambda_2^{de} \\
  & \quad - \left( 1 \leftrightarrow 2 \right) \,. \label{eq:C1402}
 \end{split}
\end{equation}
On the other hand, the commutator terms in $C_0^a\vert_{40'}$ give
\begin{equation}
 \begin{split}
  C_0^a\vert_{40'} &= V^b \Lambda_1^{de} Z^{gh,f} \left( \frac{3}{16} \epsilon_{degh[b} D_{c]f} \Lambda_2^{ac} - \frac{3}{16} \epsilon_{bcgh[d} D_{e]f} \Lambda_2^{ac} - \frac{1}{16} \epsilon_{bdghc} D_{ef} \Lambda_2^{ac} \right. \\
  & \quad \left. - \frac{1}{16} \epsilon_{beghd} D_{cf} \Lambda_2^{ac} - \frac{1}{16} \epsilon_{cdghb} D_{ef} \Lambda_2^{ac} + \frac{1}{16} \epsilon_{cdghe} D_{bf} \Lambda_2^{ac} \right) - \left( 1 \leftrightarrow 2 \right) \\
  & \quad + V^a \Lambda_1^{ac} Z^{gh,f} \left( - \frac18 \epsilon_{bdgh[c} D_{e]f} \Lambda_2^{de} + \frac18 \epsilon_{cegh[b} D_{d]f} \Lambda_2^{de} + \frac{1}{16} \epsilon_{bcgh[d} D_{e]f} \Lambda_2^{de} \right. \\
  & \quad \left. - \frac{1}{16} \epsilon_{degh[b} D_{c]f} \Lambda_2^{de} \right) - \left( 1 \leftrightarrow 2 \right) \,. \label{eq:C040}
 \end{split}
\end{equation}
Once again, we cannot have a contribution from the totally antisymmetric terms $D_{[ab} \otimes D_{cd]}$ so that \eqref{eq:C1401}, \eqref{eq:C1402} and \eqref{eq:C040} must cancel amongst themselves. To see how this works, first observe that
\begin{equation}
 \begin{split}
  \epsilon_{degh[b} Z^{gh,f} D_{c]f} &= - \epsilon_{bcgh[d} Z^{gh,f} D_{e]f} + \frac12 \epsilon_{bcdeg} Z^{hf,g} D_{hf} \,,
 \end{split}
\end{equation}
where we use, amongst other things, that $Z^{[ab,c]} = 0$. Thus, we find
\begin{equation}
 \begin{split}
  C^a\vert_{40'} &= V^b \Lambda_1^{de} Z^{gh,f} \left( \frac14 \epsilon_{bdegh} D_{cf} \Lambda_2^{ac} - \frac18 \epsilon_{cdegh} D_{bf} \Lambda_2^{ac} - \frac{3}{32} \epsilon_{bcdef} D_{gh} \Lambda_2^{ac} \right) \\
  &\quad + V^b \Lambda_1^{ac} Z^{gh,f} \left( -\frac18 \epsilon_{degh[b} D_{c]f} \Lambda_2^{de} - \frac18 \epsilon_{bdgh[c} D_{e]f} \Lambda_2^{de} + \frac18 \epsilon_{cegh[b} D_{d]f} \Lambda_2^{de} \right. \\
  & \quad \left. + \frac{1}{32} \epsilon_{bcdef} D_{gh} \Lambda_2^{de} \right) - \left( 1 \leftrightarrow 2 \right) \\
  & = V^b \Lambda_1^{de} Z^{gh,f} \left( \frac14 \epsilon_{bdegh} D_{cf} \Lambda_2^{ac} - \frac18 \epsilon_{cdegh} D_{bf} \Lambda_2^{ac} \right) + \frac18 V^b \Lambda_1^{ac} Z^{gh,f} \epsilon_{decgh} D_{bf} \Lambda_2^{de} \\
  & \quad - \frac{3}{32} V^b \Lambda_1^{de} Z^{gh,f} \epsilon_{bcdef} D_{gh} \Lambda_2^{ac} + \frac{1}{32} V^b \Lambda_1^{ac} Z^{gh,f} \epsilon_{bcdef} D_{gh} \Lambda_2^{de} - \left( 1 \leftrightarrow 2 \right) \,.
 \end{split} \label{eq:C040}
\end{equation}
Putting all this together we are left with
\begin{equation}
 \begin{split}
  C^a\vert_{40'} &= \frac{1}{16} Z^{bc,a} \epsilon_{defgh} V^d \Lambda_1^{ef} D_{bc} \Lambda_2^{gh} - \frac{3}{32} V^b \Lambda_1^{de} Z^{gh,f} \epsilon_{bcdef} D_{gh} \Lambda_2^{ac} \\
  & \quad + \frac{1}{32} V^b \Lambda_1^{ac} Z^{gh,f} \epsilon_{bcdef} D_{gh} \Lambda_2^{de} - \left( 1 \leftrightarrow 2 \right) \,. \label{eq:C40}
 \end{split}
\end{equation}
For this to vanish we require
\begin{equation}
 Z^{gh,a} D_{gh} + f^a(\mathbf{10}) = 0 \,,
\end{equation}
where $f^a(\mathbf{10})$ denotes a function of the 10-dimensional representation, which is valued in the $\mathbf{5}$ irrep.

We will now show that when this takes the form \eqref{eq:IntertwineConstraint} $C^a$ vanishes by studying the $\mathbf{10}$. First note that the totally antisymmetric terms $D_{[ab} D_{cd]}$ can now contribute. Let us take
\begin{equation}
 D_{[ab} D_{cd]} = \alpha\, \tau_{[ab} D_{cd]} \,, \qquad D_{[ab} \otimes D_{cd]} = 0 \,,
\end{equation}
and determine $\alpha$.

Let us, however, begin again with $D_1^a\vert_{10}$. Also, we will again make use of
\begin{equation}
 \tau_{bc,d}{}^a = \hat{\tau}_{bc,d}{}^a - \frac15 \delta_c^a \tau_{ab} \,, \qquad \hat{\tau}_{bc,d}{}^a = \delta^a_{[d} \tau_{bc]} \,.
\end{equation}
Then, we find
\begin{equation}
 \begin{split}
  D_1^a\vert_{10} &= \frac{1}{96} \epsilon^{abcef} \tau_{ef} \epsilon_{ghijk} V^g \Lambda_1^{hi} D_{bc} \Lambda_2^{jk} + \tau_{de} \left( \frac{1}{20} V^b \Lambda_1^{de} D_{bc} \Lambda_2^{ac} + \frac{1}{20} V^b \Lambda_1^{ac} D_{bc} \Lambda_2^{de} \right. \\
  & \quad \left. - \frac{1}{40} V^a \Lambda_1^{de} D_{bc} \Lambda_2^{bc} - \frac{1}{40} V^a \Lambda_1^{bc} D_{bc} \Lambda_2^{de} - \frac{1}{10} V^a \Lambda_1^{be} D_{bc} \Lambda_2^{cd} \right) - \left( 1 \leftrightarrow 2 \right) \\
  &= \frac{1}{96} \epsilon^{abcef} \tau_{ef} \epsilon_{ghijk} V^g \Lambda_1^{hi} D_{bc} \Lambda_2^{jk} + \tau_{de} \left( \frac{1}{20} V^b \Lambda_1^{de} D_{bc} \Lambda_2^{ac} + \frac{1}{20} V^b \Lambda_1^{ac} D_{bc} \Lambda_2^{de} \right) \\
  & - \frac{3}{20} V^a \Lambda_1^{bc} \tau_{[de} D_{bc]} \Lambda_2^{bc} - \left( 1 \leftrightarrow 2 \right) \,.
 \end{split}
\end{equation}
For $E_1^a\vert_{10}$ we find
\begin{equation}
 \begin{split}
  E_1^a\vert_{10} &= V^b \Lambda_1^{de} \left( \frac16 \tau_{ef} D_{bd} \Lambda_2^{af} + \frac13 \tau_{be} D_{cd} \Lambda_2^{ac} + \frac15 \tau_{de} D_{bc} \Lambda_2^{ac} \right) \\
  & \quad + V^b \left( - \frac{2}{15} \Lambda_1^{ac} \tau_{de} D_{bc} \Lambda_2^{de} + \frac16 \Lambda_1^{ad} \tau_{de} D_{bc} \Lambda_2^{ce} \right) - \left( 1 \leftrightarrow 2 \right) \,.
 \end{split}
\end{equation}
On the other hand from $C_0^a\vert_{10}$ we obtain
\begin{equation}
 \begin{split}
  C_0^a\vert_{10} &= V^b \Lambda_1^{ac} \left( \frac1{12} \tau_{bd} D_{ce} \Lambda_2^{de} - \frac1{12} \tau_{ce} D_{bd} \Lambda_2^{de} - \frac1{24} \tau_{bc} D_{de} \Lambda_2^{de} + \frac1{24} \tau_{de} D_{bc} \Lambda_2^{de}  \right) \\
  & \quad + V^b \Lambda_1^{de} \left( + \frac18 \tau_{bc} D_{de} \Lambda_2^{ac} -\frac18 \tau_{de} D_{bc} \Lambda_2^{ac} + \frac1{12} \tau_{bd} D_{ce} \Lambda_2^{ac} + \frac1{12} \tau_{cd} D_{be} \Lambda_2^{ac} \right) \\
  & \quad - \frac{3}{4}\alpha V^b \Lambda_1^{de} \tau_{[bc} D_{de]} \Lambda_2^{ac} - \frac34 \alpha V^b \Lambda_1^{ac} \tau_{[bc} D_{de]} \Lambda_2^{de} + \frac3{10} \alpha V^a \Lambda_1^{de} \tau_{[de} D_{bc]} \Lambda_2^{bc} \\
  & \quad - \left( 1 \leftrightarrow 2 \right) \\
  &= V^b \Lambda_1^{ac} \left( - \frac14  \tau_{[bc} D_{de]} \Lambda_2^{de} - \frac16 \tau_{ce} D_{bd} \Lambda_2^{de} + \frac1{12} \tau_{de} D_{bc} \Lambda_2^{de} \right) \\
  & \quad V^b \Lambda_2^{de} \left( \frac34 \tau_{[de} D_{bc]} \Lambda_2^{ac} - \frac14 \tau_{de} D_{bc} \Lambda_2^{ac} - \frac16 \tau_{cd} D_{be} \Lambda_2^{ac} + \frac13 \tau_{bd} D_{ce} \Lambda_2^{ac} \right) \\
  & \quad - \frac{3}{4}\alpha V^b \Lambda_1^{de} \tau_{[bc} D_{de]} \Lambda_2^{ac} - \frac34 \alpha V^b \Lambda_1^{ac} \tau_{[bc} D_{de]} \Lambda_2^{de} + \frac3{10} \alpha V^a \Lambda_1^{de} \tau_{[de} D_{bc]} \Lambda_2^{bc} \\
  & \quad - \left( 1 \leftrightarrow 2 \right) \,.
 \end{split}
\end{equation}
Putting all this together we find
\begin{equation}
 \begin{split}
  C^a\vert_{10} &= \frac1{96} \epsilon^{abcij} \tau_{ij} \epsilon_{defgh} V^d \Lambda_1^{ef} D_{bc} \Lambda_2^{gh} - \frac14 V^b \Lambda_1^{ac} \tau_{[bc} D_{de]} \Lambda_2^{de} + \frac34 V^b \Lambda_1^{de} \tau_{[bc} D_{de]} \Lambda_2^{ac} \\
  & \quad - \frac3{20} V^a \Lambda_1^{bc} \tau_{[de} D_{bc]} \Lambda_2^{bc} + \frac3{10} \alpha V^a \Lambda_1^{de} \tau_{[de} D_{bc]} \Lambda_2^{bc} - \frac34 \alpha V^b \Lambda_1^{de} \tau_{[bc} D_{de]} \Lambda_2^{ac} \\
  & \quad - \frac34 \alpha V^b \Lambda_1^{ac} \tau_{[bc} D_{de]} \Lambda_2^{de} - \left( 1 \leftrightarrow 2 \right) \\
  &= \frac1{96} \epsilon^{abcij} \tau_{ij} \epsilon_{defgh} V^d \Lambda_1^{ef} D_{bc} \Lambda_2^{gh} + \frac54 V^b \Lambda_1^{ac} \tau_{[de} D_{bc]} \Lambda_2^{de} + \frac94 V^b \Lambda_1^{de} \tau_{[bc} D_{de]} \Lambda_2^{ac} \\
  & \quad - \frac34 V^a \Lambda_1^{bc} \tau_{[de} D_{bc]} \Lambda_2^{bc} - \left(\alpha+2\right) \left( \frac34 V^b \Lambda_1^{de} \tau_{[bc} D_{de]} \Lambda_2^{ac} + \frac34 V^b \Lambda_1^{ac} \tau_{[bc} D_{de]} \Lambda_2^{de} \right. \\
  & \quad \left. - \frac3{10} V^a \Lambda_1^{de} \tau_{[de} D_{bc]} \Lambda_2^{bc} \right) - \left( 1 \leftrightarrow 2 \right) \\
  &= \frac1{96} \epsilon^{abcij} \tau_{ij} \epsilon_{defgh} V^d \Lambda_1^{ef} D_{bc} \Lambda_2^{gh} - \frac{5!}{32} \tau_{de} V^{[a} \Lambda_1^{bc} D_{bc} \Lambda_2^{de]} + \frac{4!}{32} V^b \Lambda_1^{de} \tau_{[de} D_{bc]} \Lambda_2^{ac} \\
  & \quad - \frac{4!}{96} V^b \Lambda_1^{ac} \tau_{[de} D_{bc]} \Lambda_2^{de} - \left(\alpha+2\right) \left( \frac34 V^b \Lambda_1^{de} \tau_{[bc} D_{de]} \Lambda_2^{ac} + \frac34 V^b \Lambda_1^{ac} \tau_{[bc} D_{de]} \Lambda_2^{de} \right. \\
  & \quad \left. - \frac3{10} V^a \Lambda_1^{de} \tau_{[de} D_{bc]} \Lambda_2^{bc} \right) - \left( 1 \leftrightarrow 2 \right) \,.
 \end{split}
\end{equation}
We want to combine these different terms into expressions involving contractions of two $\epsilon_{abcde}$ symbols, to make contact with \eqref{eq:C40}. We find
\begin{equation}
 \begin{split}
  C^a\vert_{10} &= \frac1{96} \epsilon^{abcij} \tau_{ij} \epsilon_{defgh} V^d \Lambda_1^{ef} D_{bc} \Lambda_2^{gh} - \frac{1}{32} \epsilon^{abcij} \tau_{ij} \epsilon_{defgh} V^d \Lambda_1^{ef} D_{bc} \Lambda_2^{gh} \\
  & \quad + \frac{1}{32} \epsilon^{fghij} \epsilon_{bcdef} V^b \Lambda_1^{de} \tau_{ij} D_{gh} \Lambda_2^{ac} - \frac{1}{96} \epsilon^{fghij} \epsilon_{bcdef} V^b \Lambda_1^{ac} \tau_{ij} D_{gh} \Lambda_2^{de} \\
  & \quad - \left(\alpha+2\right) \left( \frac34 V^b \Lambda_1^{de} \tau_{[bc} D_{de]} \Lambda_2^{ac} + \frac34 V^b \Lambda_1^{ac} \tau_{[bc} D_{de]} \Lambda_2^{de} \right. \\
  & \quad \left. - \frac3{10} V^a \Lambda_1^{de} \tau_{[de} D_{bc]} \Lambda_2^{bc} \right) - \left( 1 \leftrightarrow 2 \right) \\
  &= - \frac{1}{48} \epsilon^{abcij} \tau_{ij} \epsilon_{defgh} V^d \Lambda_1^{ef} D_{bc} \Lambda_2^{gh} + \frac{1}{32} \epsilon^{fghij} \epsilon_{bcdef} V^b \Lambda_1^{de} \tau_{ij} D_{gh} \Lambda_2^{ac} \\
  & \quad - \frac{1}{96} \epsilon^{fghij} \epsilon_{bcdef} V^b \Lambda_1^{ac} \tau_{ij} D_{gh} \Lambda_2^{de} - \left(\alpha+2\right) \left( \frac34 V^b \Lambda_1^{de} \tau_{[bc} D_{de]} \Lambda_2^{ac} \right. \\
  & \quad \left. + \frac34 V^b \Lambda_1^{ac} \tau_{[bc} D_{de]} \Lambda_2^{de} - \frac3{10} V^a \Lambda_1^{de} \tau_{[de} D_{bc]} \Lambda_2^{bc} \right) - \left( 1 \leftrightarrow 2 \right)
 \end{split} \label{eq:C10}
\end{equation}
Thus, we have in total
\begin{equation}
 \begin{split}
  C^a_0 + C^a_1 &= \frac{1}{16} \left( Z^{bc,a} - \frac13 \epsilon^{abcij} \tau_{ij} \right) \epsilon_{defgh} V^d \Lambda_1^{ef} D_{bc} \Lambda_2^{gh} \\
  & \quad - \frac{3}{32} V^b \Lambda_1^{de} \left( Z^{gh,f} - \frac13 \epsilon^{fghij} \tau_{ij} \right) \epsilon_{bcdef} D_{gh} \Lambda_2^{ac} \\
  & \quad + \frac{1}{32} V^b \Lambda_1^{ac} \left( Z^{gh,f} - \frac13 \epsilon^{fghij} \tau_{ij} \right) \epsilon_{bcdef} D_{gh} \Lambda_2^{de} \\
  & \quad - \left(\alpha+2\right) \left( \frac34 V^b \Lambda_1^{de} \tau_{[bc} D_{de]} \Lambda_2^{ac} + \frac34 V^b \Lambda_1^{ac} \tau_{[bc} D_{de]} \Lambda_2^{de} \right. \\
  & \quad \left. - \frac3{10} V^a \Lambda_1^{de} \tau_{[de} D_{bc]} \Lambda_2^{bc} \right) - \left( 1 \leftrightarrow 2 \right) \,,
 \end{split}
\end{equation}
which vanishes if we impose $\alpha = - 2$ as in \eqref{eq:SecCon} and the constraint \eqref{eq:IntertwineConstraint}
\begin{equation}
 \left( Z^{ab,c} - \frac13 \epsilon^{abcde} \tau_{de} \right) D_{ab} = 0 \,. \label{eq:ZD0}
\end{equation}
Note that this is equivalent to the symmetric part of the hatted embedding tensor $\hat{\tau}_{ab,c}{}^d$ in the $\mathbf{10}$-representation vanishing:
\begin{equation}
 \hat{\tau}_{ab,cd}{}^{ef} + \hat{\tau}_{cd,ab}{}^{ef} = 0 \,,
\end{equation}
and hence we can also rewrite the commutator condition \eqref{eq:ACommRel} as
\begin{equation}
 \left[ D_{bc}, D_{de} \right] = 2 \tau_{bc,[d}{}^{[f} \delta_{e]}{}^{g]} + \frac25 \tau_{bc} \delta_{de}^{fg} \,. \label{eq:ACommRelFin}
\end{equation}

\section{Curvature}\label{A:Curvature}
The curvature,
\begin{equation}
 R_{MN,a}{}^b = \frac14 E^{cd}{}_M E^{ef}{}_N R_{cd,ef,a}{}^b \,,
\end{equation}
is traceless and thus lives in the irreducible representations
\begin{equation}
 \mathbf{45} \otimes \mathbf{24} = \mathbf{5} \oplus \mathbf{45} \oplus \mathbf{45} \oplus \mathbf{50} \oplus \mathbf{70} \oplus \mathbf{105} \oplus \mathbf{280} \oplus \mathbf{480} \,.
\end{equation}
Note first of all that the $\mathbf{45}$ can be described by the antisymmetric product of $\mathbf{10}$'s or equivalently as the traceless product of $\mathbf{5} \otimes \overline{\mathbf{10}}$. Let us therefore define
\begin{equation}
 \left( R_{a}{}^{bc} \right)^d{}_e = \frac{1}{3!} \epsilon^{bcfgh} R_{af,gh,e}{}^d \,.
\end{equation}
This can be inverted as follows
\begin{equation}
 R_{ab,cd,e}{}^f = \frac32 \epsilon_{cdkl[b} A_{a]}{}^{kl} \,.
\end{equation}
Let us now give the different irreducible components of the curvature:\\
$\mathbf{5}$
\begin{equation}
 ( R_{b}{}^{ac} )^b{}_c = \frac{7}{8} \left( \frac14 S_{bc} Z^{ab,c} - \frac{19}{27} \tau_{bc} Z^{bc,a} - \frac{40}{567} \epsilon^{abcde} \tau_{bc} \tau_{de} \right) \,.
\end{equation}
Using the quadratic constraints \eqref{eq:QCLin} we can write this as
\begin{equation}
 ( R_{b}{}^{ac} )^b{}_c = \frac{221}{1296} \epsilon^{abcde} \tau_{bc} \tau_{de} \,.
\end{equation}
The remaining irreducible representations are, up to the quadratic constraint, as follows. The $\mathbf{70}$ and the two $\mathbf{45}$'s are given by\\
$\mathbf{70}$
\begin{equation}
 \begin{split}
  (R_{a}{}^{d(b})^{c)}{}_d + \frac{1}{6} \delta^{(c}_a (R_e{}^{b)f})^{e}{}_f &= -\frac{35}{864} S_{ad} Z^{d(b,c)} - \frac{35}{2592} \delta_a^{(b} \epsilon^{c)defg} \tau_{de} \tau_{fg} + \frac{85}{1296} \epsilon_{adefg} Z^{de,b} Z^{fg,c} \,.
 \end{split}
\end{equation}
$\mathbf{45}_1$
\begin{equation}
 \begin{split}
  (R_{a}{}^{d[b})^{c]}{}_d ) - \frac14 \delta_a^{[b} (R_d{}^{c]e})^d{}_e &= - \frac7{384} \epsilon^{bcdef} S_{ad} \tau_{ef} \,.
 \end{split}
\end{equation}
$\mathbf{45}_2$
\begin{equation}
 \begin{split}
  ( R_d{}^{bc})^{d]}{}_a + \frac12 \delta_a^{[b} (R_d{}^{c]e})^{d}{}_e &= \frac{67}{1728} \epsilon^{bcdef} S_{ad} \tau_{ef} \,.
 \end{split}
\end{equation}
For the $\mathbf{50}$ consider\\
\begin{equation}
 \hat{R}_{ab,cd} = R_{a[c,de],b}{}^e
\end{equation}
and
\begin{equation}
 \tilde{R}_{ab,cd} = \hat{R}_{ab,cd} - \hat{R}_{ba,cd} + \hat{R}_{cd,ab} - \hat{R}_{dc,ab} \,,
\end{equation}
Then it is given by
\begin{equation}
 \begin{split}
  \tilde{R}_{ab,cd} - \tilde{R}_{a[b,cd]} &= \frac3{16} S_{a[c} S_{d]b} - \frac{220}{81} \tau_{a[c} \tau_{d]b} - \frac{4}{27} \epsilon_{abefg} \epsilon_{cdhij} Z^{ef,h} Z^{ij,g} + \frac{220}{81} \tau_{a[b} \tau_{cd]} \,.
 \end{split}
\end{equation}
The $\mathbf{105}$ gives rise to\\
\begin{equation}
 \begin{split}
  \left(P_{105} \hat{R}\right)_{ab,cd} &= P_{105}\left(-\frac{1}{24} S_{ab} \tau_{cd} - \frac{1}{192} S_{e(b} \epsilon_{a)cdfg} Z^{fg,e} - \frac{23}{216} \tau_{e(b} \epsilon_{a)cdfg} Z^{fg,e} \right) \\
  &= -\frac{1}{24} S_{ab} \tau_{cd} - \frac{1}{192} S_{e(b} \epsilon_{a)cdfg} Z^{fg,e} - \frac{23}{216} \tau_{e(b} \epsilon_{a)cdfg} Z^{fg,e} + \ldots \,,
 \end{split}
\end{equation}
where $P_{105}$ denotes the projector onto this representation and the $\ldots$ refer to contractions and (anti-)symmetrisations. For the $\mathbf{280}$ we obtain
\begin{equation}
 \begin{split}
  (P_{280}R)_{ab}{}^{cd,e} &= P_{280}\left( - \frac{1}{81} \tau_{ab} Z^{cd,e} - \frac{19}{216} \epsilon_{abfgh} Z^{cd,f} Z^{gh,e} + \frac{1}{36} \epsilon_{abfgh} Z^{gh,[c} Z^{d]f,e} \right) \\
  &= - \frac{1}{81} \tau_{ab} Z^{cd,e} - \frac{19}{216} \epsilon_{abfgh} Z^{cd,f} Z^{gh,e} + \frac{1}{36} \epsilon_{abfgh} Z^{gh,[c} Z^{d]f,e} + \ldots \,,
 \end{split}
\end{equation}
where $P_{280}$ denotes the projector onto the $\mathbf{280}$ and the $\ldots$ refer to contractions and \\(anti-)symmetrisations. Finally, there is the $\mathbf{480}$
\begin{equation}
 (P_{480}R)_{ab}{}^{cd,e} = P_{480} \left( - \frac{5}{288} S_{ab} Z^{cd,e} \right) = - \frac5{288} S_{ab} Z^{cd,e} + \ldots \,,
\end{equation}
where $P_{480}$ denotes the projector onto the $\mathbf{480}$ and the $\ldots$ refer to contractions.

\section{Scalar potential}\label{A:ScalarPot}
\subsection{Lorentz-invariance vs diffeomorphism invariance}
Here we will show the details that allow one to rewrite the action \eqref{eq:V1EmbTensor} in terms of the generalised metric as in \eqref{eq:MetricPotential}.

Let us first of all rewrite the terms in \eqref{eq:V1EmbTensor} in terms of the fluctuation connection $\tGamma$.
\begin{equation}
 \begin{split}
  A &= 8 \gM^{ac} \gM^{bd} \left( \tGamma_{ea,b}{}^e \tGamma_{fc,d}{}^f + \tGamma_{ea,b}{}^e \tGamma_{fd,c}^{}f \right) \,, \\
  B &= 16 \gM^{ac} \gM^{bd} \tGamma_{ea,c}{}^e \tGamma_{fb,d}{}^f \,, \\
  C &= \gM^{ac} \gM^{bd} \left( \frac{36}{25} \gamma_{ab} \gamma_{cd} + \frac{24}{5} \gamma_{ab} \tGamma_{ec,d}{}^e + 2 \tGamma_{ea,b}{}^e \tGamma_{fc,d}{}^f - \tGamma_{ea,b}{}^e \tGamma_{fd,c}{}^f \right) \,, \\
  D &= \gM^{ac} \gM^{bd} \left( 2 \tGamma_{ae,b}{}^f \tGamma_{cf,d}{}^e - \frac23 \tGamma_{ae,b}{}^e \tGamma_{cf,d}{}^f + \frac23 \tGamma_{ae,b}{}^e \tGamma_{df,c}{}^f + \gM^{ef} \gM_{gh} \tGamma_{ab,e}{}^g \tGamma_{cd,f}{}^h \right) \,,\\
  E &= \gM^{ac} \gM^{bd} \left( - \frac12 \tGamma_{ab,e}{}^f \tGamma_{cd,f}{}^e + 2 \tGamma_{ab,e}{}^f \tGamma_{cf,d}{}^e + \tGamma_{ae,b}{}^f \tGamma_{df,c}{}^e + \frac12 \gM^{ef} \gM_{gh} \tGamma_{ab,e}{}^g \tGamma_{cd,f}{}^h \right) \,, \\
  F &= \gM^{ac} \gM^{bd} \left( 2 \nabla_{ab} \tGamma_{ec,d}{}^e + \frac65 \nabla_{ab} \gamma_{cd} + \frac{6}{25} \gamma_{ab} \gamma_{cd} + \frac25 \gamma_{ab} \tGamma_{ec,d}{}^e - \frac{12}{5} \tGamma_{ab,c}{}^e \gamma_{ed} \right. \\
  & \left. \quad - 2 \tGamma_{ab,c}{}^e \tGamma_{fe,d}{}^f - \tGamma_{ab,c}{}^e \tGamma_{df,e}{}^e \right) \,.
 \end{split}
\end{equation}

Let us compare this to the possible terms of the action which involve two derivatives of the generalised metric. These are given in terms of the fluctuation connections as
\begin{equation}
 \begin{split}
  \gM^{ac} \gM^{bd} \nabla_{ab} \gM^{ef} \nabla_{cd} \gM_{ef} &= -2 \gM^{ac} \gM^{bd} \tGamma_{ab,f}{}^e \tGamma_{cd,e}{}^f - 2 \gM^{ac} \gM^{bd} \gM^{ef} \gM_{gh} \tGamma_{ab,e}{}^g \tGamma_{cd,f}{}^h \,, \\
  \gM^{ac} \gM^{bd} \nabla_{ab} \gM^{ef} \nabla_{ec} \gM_{df} &= 2 \gM^{ac} \gM^{bd} \tGamma_{ab,e}{}^f \tGamma_{cf,d}{}^e \,, \\
  \nabla_{ab} \gM^{ac} \nabla_{cd} \gM^{bd} &= \gM^{ac} \gM^{bd} \left( \tGamma_{ef,c}{}^e \tGamma_{ab,d}{}^f + \tGamma_{ae,c}{}^f \tGamma_{fb,d}{}^e + \tGamma_{eb,a}{}^e \tGamma_{cf,d}{}^f + \tGamma_{ab,c}{}^e \tGamma_{ef,d}{}^f \right) \,, \\
  \gM^{ac} \nabla_{ab} \nabla_{cd} \gM^{bd} &= \gM^{ac} \gM^{bd} \left( -2 \nabla_{ab} \tGamma_{ce,d}{}^e + 2 \tGamma_{ab,d}{}^e \tGamma_{cf,e}{}^f + \tGamma_{ae,b}{}^e \tGamma_{cf,d}{}^f \right. \\
  & \quad \left. + \tGamma_{ae,b}{}^f \tGamma_{cf,d}{}^e - \tE_{\bee}{}^e \left[ \nabla_{ae} ,\, \nabla_{cd} \right] \tE_{b}{}^{\bee} \right) \,. \label{eq:MetricExpansion}
 \end{split}
\end{equation}
Let us explain how the last equation comes about in more detail. We use the fact that
\begin{equation}
 \begin{split}
  \gM^{ac} \nabla_{ab} \nabla_{cd} \gM^{bd} &= \gM^{ac} \nabla_{ab} \left( \tGamma_{cd,e}{}^b \gM^{ed} + \tGamma_{cd,e}{}^d \gM^{be} \right) \\
  & = \gM^{ac} \gM^{bd} \left( \nabla_{ae} \tGamma_{cd,b}{}^e + \nabla_{cd} \tGamma_{ce,b}{}^e \right) + \ldots \,, \label{eq:DoubleDerivatives}
 \end{split}
\end{equation}
where the $\ldots$ denote $\tGamma^2$ terms. We also have
\begin{equation}
 \tnabla_{ab} \tGamma_{cd,f}{}^e - \tnabla_{cd} \tGamma_{ab,f}{}^e = \tGamma_{ab,f}{}^g \tGamma_{cd,g}{}^e - \tGamma_{ab,g}{}^e \tGamma_{cd,f}{}^g + \tE_{\bee}{}^e \left[ \tnabla_{ab},\, \tnabla_{cd} \right] \tE_{f}{}^{\bee} \,, \label{eq:FlatConnection}
\end{equation}
which is analogous to the statement that the usual Weitzenb\"ock connection is flat. Using \eqref{eq:FlatConnection} we can write \eqref{eq:DoubleDerivatives} as in \eqref{eq:MetricExpansion}.

One can now check that it is possible to write the potential \eqref{eq:V1EmbTensor} in terms of \eqref{eq:MetricExpansion} as follows:
\begin{dmath}
 S_{pot} = \int \IM \Bigg( \frac18 \gM^{ac} \gM^{bd} \nabla_{ab} g^{\mu\nu} \nabla_{cd} g_{\mu\nu} - \frac{5}{14} \gM^{ac} \gM^{bd} D_{ab} \ln e \, D_{cd} \ln e - \frac{12}{7} \gM^{ac} D_{ab} \gM^{bd} D_{cd} \ln e - \frac67 \gM^{ac} \gM^{bd} \nabla_{ab} D_{cd} \ln e - \frac12 \nabla_{ab} \gM^{ac} \nabla_{cd} \gM^{bd} + \frac18 \gM^{ac} \gM^{bd} \nabla_{ab} \gM^{ef} \nabla_{cd} \gM_{ef} + \frac12 \gM^{ac} \gM^{bd} \nabla_{ab} \gM^{ef} \nabla_{ec} \gM_{df} - \gM^{ac} \nabla_{ab} \nabla_{cd} \gM^{bd} + \Delta V \Bigg) \,. \label{eq:MetricPotential}
\end{dmath}
We write the anomalous terms as $\Delta V = \Delta + \delta$ with the individual pieces given by
\begin{equation}
 \begin{split}
  \Delta &= \frac12 \gM^{ac} \gM^{bd} \left( - \tGamma_{ae,b}{}^e \tGamma_{df,c}{}^f - 2 \tGamma_{ab,c}{}^e \tGamma_{ef,d}{}^f + \tGamma_{ae,c}{}^e \tGamma_{bf,d}{}^f + \tGamma_{af,b}{}^e \tGamma_{de,c}{}^f - \tGamma_{af,c}{}^e \tGamma_{be,d}{}^f \right) \,, \\
  \delta &= \gM^{ac} \gM^{bd} \tE_{\bee}{}^e \left[ \tnabla_{ab},\, \tnabla_{cd} \right] \tE^{\bee}{}_f \\
  &= \gM^{ac} \tE^{\bbb b} \left( 2 \tau_{ae,[b}{}^f D_{c]f} \tE_{\bbb}{}^e - \frac45 \tau_{ae} D_{bc} \tE_{\bbb}{}^e - 2 \Conn_{ae,[b}{}^f \nabla_{c]f} \tE_{\bbb}{}^e + 2 \Conn_{bc,[a}{}^f \nabla_{e]f} \tE_{\bbb}{}^e \right) \,.
 \end{split}
\end{equation}
To analyse these two terms further, let us split
\begin{equation}
 \tGamma_{ab,c}{}^d = \Gammao_{ab,c}{}^d - \Conn_{ab,e}{}^c\, \gE^{\bbc}{}_c \gE_{\bbd}{}^e \,,
\end{equation}
where
\begin{equation}
 \Gammao_{ab,c}{}^d = - \gE^{\bbc}{}_c D_{ab} \gE_{\bbd}{}^c \,.
\end{equation}
Let us now expand $\Delta$ in terms of pieces independent of $\Conn$, labelled $\Delta_0$, those linear in $\Conn$, labelled $\Delta_1$, and those quadratic in $\Conn$, labelled $\Delta_2$. Similarly, $\delta$ has terms linear in $\tau$ and $\Conn$, labelled $\delta_1$, and terms quadratic in $\Conn$, labelled $\delta_2$. We find that $\Delta_0$ is given by
\begin{equation}
 \Delta_0 = \frac12 \gM^{ac} \gM^{bd} \left( - \Gammao_{ae,b}{}^e \Gammao_{df,c}{}^f - 2 \Gammao_{ab,c}{}^e \Gammao_{ef,d}{}^f + \Gammao_{ae,c}{}^e \Gammao_{bf,d}{}^f + \Gammao_{af,b}{}^e \Gammao_{de,c}{}^f - \Gammao_{af,c}{}^e \Gammao_{be,d}{}^f \right) \,, \label{eq:ConnSC}
\end{equation}
and vanishes by the section condition. This is exactly the piece which is required in the gauged EFT set-up to rewrite the action as a function of the embedding tensor in terms of the generalised metric \cite{Blair:2014zba}.

The terms in $\Delta_1$ and $\delta_1$ are not $\SO{5}$-invariant and thus they need to cancel. Using the linear constraint we find indeed that for vanishing background trombone $\tau_{ab} = 0$ their contributions cancel, specifically
\begin{equation}
 \Delta_1 = - \delta_1 = \gM^{ac} \tE^{b\bbb} \left( \frac18 \tS_{ab} D_{ce} \tE_{\bbb}{}^e - \frac18 \tS_{ce} D_{ab} \tE_{\bbb}{}^e - \frac18 \tS_{ac} D_{be} \tE_{\bbb}{}^e + \frac16 \epsilon_{abegh} \tZ^{gh,f} D_{cf} \tE_{\bbb}{}^e \right) \,.
\end{equation}
We are left with the terms quadratic in $\Conn$ given by
\begin{dmath}
 \Delta_2 + \delta_2 =  \gM^{ac} \gM^{bd} \bigg( - \frac12 \Conn_{ae,c}{}^e \Conn_{bf,d}{}^f - \frac12 \Conn_{ae,c}{}^f \Conn_{bf,d}{}^e + \frac12 \Conn_{ae,b}{}^e \Conn_{df,c}{}^f - \frac12 \Conn_{ae,b}{}^f \Conn_{df,c}{}^e - \Conn_{ae,d}{}^f \Conn_{fc,b}{}^e - \Conn_{ae,f}{}^e \Conn_{cd,b}{}^f \bigg) \,.
\end{dmath}

\subsection{Rewriting using ``big'' generalised metric}\label{A:BigMetric}
Let us first fix our notation. We label antisymmetric pairs of indices as $A = [ab]$, $A = 1, \ldots, 10$. To avoid double counting every contraction of these 10-d indices $A, B$ comes with a factor of $\frac12$ when written in terms of the fundamental $\SLf$ indices. For example, we would write
\begin{equation}
 V^A W_A = \frac12 V^{ab} W_{ab} \,.
\end{equation}
Now consider the possible terms which involve two derivatives of the generalised metric. For simplicity, we will work with the first order action so there are no total derivative terms. Then we can have the following terms:
\begin{equation}
 \begin{split}
  M^{AB} \nabla_{A} M^{CD} \nabla_{B} M_{CD} &= \frac{1}{16} M^{ab,cd} \nabla_{ab} M^{ef,gh} \nabla_{cd} M_{ef,gh} \\
  &= \frac{3}{2} \gM^{ac} \gM^{bd} \nabla_{ab} \gM^{ef} \nabla_{cd} \gM_{ef} \,, \\
  M^{AB} \nabla_A M^{CD} \nabla_C M_{BD} &= \frac{1}{16} M^{ab,cd} \nabla_{ab} M^{ef,gh} \nabla_{ef} M_{cd,gh} \\
  &= - \gM^{ac} \gM^{bd} \nabla_{ab} \gM^{ef} \nabla_{ec} \gM_{df} - \nabla_{ab} \gM^{ac} \nabla_{cd} \gM^{bd} \,, \\
  \nabla_A M^{AB} \nabla_B \ln g &= \frac14 \nabla_{ab} M^{ab,cd} \nabla_{cd} \ln g \\
  &= \gM^{ac} \nabla_{ab} \gM^{bd} \nabla_{cd} \ln g \,.
 \end{split}
\end{equation}
These allow us to rewrite the terms which do not contain $\Conn^2$ explicitly as
\begin{equation}
 \begin{split}
  S_{pot} &= \int d^{10}Y d^7x \, \gE e \left( \frac14 M^{AB} \nabla_{A} g^{\mu\nu} \nabla_{B} g_{\mu\nu} + \frac14 M^{AB} D_{A} \ln g \, D_{B} \ln g + \frac12 \nabla_A M^{AB} \nabla_B \ln g  \right. \\
  & \left. \quad + \frac{1}{12} M^{AB} \nabla_A M^{CD} \nabla_B M_{CD} - \frac{1}{2} M^{AB} \nabla_A M^{CD} \nabla_{C} M_{BD} \right) + \ldots \,.
 \end{split}
\end{equation}

Finally, let us look at the terms involving $\Conn^2$. To this end we consider
\begin{equation}
 \begin{split}
  M^{ab,cd} \Conn_{ae,c}{}^f \Conn_{bf,d}{}^e &= \gM^{ac} \gM^{bd} \left( \Conn_{ae,c}{}^f \Conn_{bf,d}{}^e - \Conn_{ae,d}{}^f \Conn_{bf,c}{}^e \right) \,, \\
  M^{ab,cd} \Conn_{ae,c}{}^e \Conn_{bf,d}{}^f &= \gM^{ac} \gM^{bd} \left( \Conn_{ae,c}{}^e \Conn_{bf,d}{}^f - \Conn_{ae,b}{}^e \Conn_{df,c}{}^f \right) \,, \\
  M^{ab,cd} \Conn_{ae,d}{}^f \Conn_{fc,b}{}^e &= \gM^{ac} \gM^{bd} \left( \Conn_{ae,d}{}^f \Conn_{fc,b}{}^e - \Conn_{ae,c}{}^f \Conn_{fd,b}{}^e \right) \,, \\
  M^{ab,cd} \Conn_{ae,f}{}^e \Conn_{cd,b}{}^f &= 2 \gM^{ac} \gM^{bd} \Conn_{ae,f}{}^e \Conn_{cd,b}{}^f \,,
 \end{split}
\end{equation}
and find that we can write the $\Conn^2$ terms as
\begin{equation}
 \begin{split}
  S_{pot} &= \int d^{10}Y d^7x \, \gE e M^{ab,cd} \left( -\frac12 \Conn_{ae,c}{}^e \Conn_{bf,d}{}^f - \frac12\Conn_{ae,f}{}^e \Conn_{cd,b}{}^f - \frac12\Conn_{ae,c}{}^f \Conn_{bf,d}{}^e + \Conn_{ae,d}{}^f \Conn_{fc,b}{}^e \right) \\
  & \quad + \ldots \,.
 \end{split}
\end{equation}

\section{External Diffeomorphisms}\label{A:ExtDiffeos}
\subsection{Topological term and gauge kinetic terms}
To begin let us split the variation of the gauge field $\delta_\xi \Aa_\mu$ into $\delta_\xi^0 A_\mu$, which does not depend on $\gM_{ab}$, and $\delta_\xi^1 A_\mu$, which does. Thus, the relation
\begin{equation}
 \delta_\xi \Aa_\mu{}^{ab} = \delta^{0}_\xi \Aa_\mu{}^{ab} + \delta^{1}_\xi \Aa_{\mu}{}^{ab}\,,
\end{equation}
with
\begin{equation}
 \delta^{0}_\xi \Aa_\mu{}^{ab} = \xi^\nu \Fa_{\nu\mu}{}^{ab} \,, \qquad \delta^{1}_\xi \Aa_{\mu}{}^{ab} = \gM^{ac} \gM^{bd} g_{\mu\nu} \nabla_{cd} \xi^\nu \,,
\end{equation}
holds. Now the $\delta_\xi^0$ variation of the field strengths is given by
\begin{equation}
 \begin{split}
  \delta_\xi^0 \Fa_{\mu\nu}{}^{ab} &= L_{\xi} \Fa_{\mu\nu}{}^{ab} + \frac{1}{2} \epsilon^{abcde} \Fb_{\mu\nu\rho,c} \nabla_{de} \xi^{\rho} \,, \\
  \delta_\xi^0 \Fb_{\mu\nu\rho,a} &= L_\xi \Fb_{\mu\nu\rho\,a} + \Fc_{\sigma\mu\nu\rho}{}^{b} \nabla_{ba} \xi^{\sigma} \,,
 \end{split}
\end{equation}
where $L_\xi$ is exterior Lie derivative which is the usual 7-dimensional Lie derivative but with the covariant derivative $\D_\mu$. We will also need
\begin{equation}
 \delta_\xi^1 \Fb_{\mu\nu\rho,a} = - \frac34 \epsilon_{abcde} \gM^{bf} \gM^{cg} \nabla_{fg} \xi^\sigma g_{\sigma[\mu} \Fa_{\nu\rho]}{}^{de} \,.
\end{equation}
Using these results, we find the anomalous variation of the topological term to take a simple form. Here we express it as a variation of the 7-dimensional Lagrangian,
\begin{equation}
 \begin{split}
 \Delta_\xi L_{top} &= - \frac{1}{\sqrt{6}} \left( 2 \gM^{ac} {\gM}^{bd} \nabla_{cd} \xi^\nu g_{\mu_1\nu} \Fc_{\mu_2\mu_3\mu_4,c} \Fc_{\mu_5\mu_6\mu_6,d} + \nabla_{ab} \left( \xi^\nu \Fc_{\nu\mu_1\ldots\mu_3}{}^a \right) \Fc_{\mu_4\ldots\mu_7}{}^b \right) \,. \label{eq:deltaTop}
 \end{split}
\end{equation}

Let us now turn to the gauge-kinetic term $S_{GK}$. It is easy to see that the $\delta^0$ variation of the $\Fa^2$ term cancels the $\delta^1$ variation of the $\Fb^2$ term when the relative coefficients are exactly $\frac23$. The $\delta^1$ variation of $\Fa^2$ will be used to cancel against the variation of the other terms and we will return to this later. For now, let us consider the $\delta^0$ variation of $\Fb^2$. It is given by
\begin{equation}
 \delta_\xi^0 L_{GK} = - \frac16 e \gM^{ab} \Fb^{\mu\nu\rho}{}_a \nabla_{cb} \xi^\sigma \Fc_{\sigma\mu\nu\rho}{}^c \,. \label{eq:delta0GK}
\end{equation}
This turns out to cancel the variation of the topological term \eqref{eq:deltaTop}, up to a self-duality equation that comes from the equations of motion. Consider the variation of $\Delta \Ac_{\mu\nu\rho}$. The associated equations of motion are
\begin{equation}
 \nabla_{ca} \left( \frac{1}{2\sqrt{6}} \epsilon^{\mu_1\ldots\mu_7} \Fc_{\mu_4\ldots\mu_7}{}^a - \frac{1}{12} e \Fb^{\mu_1\mu_2\mu_3}{}_b {\cal M}^{ab} \right) = 0 \,.
\end{equation}
As usual \cite{Hohm:2013vpa}, we take this projected duality equation to hold outside the derivative too, so that
\begin{equation}
 \frac{1}{2\sqrt{6}} \epsilon^{\mu_1\ldots\mu_7} \Fc_{\mu_4\ldots\mu_7}{}^a = \frac{1}{12} e \Fb^{\mu_1\mu_2\mu_3}{}_b {\cal M}^{ab} \,.
\end{equation}
It is now easy to see that the variations \eqref{eq:delta0GK} and \eqref{eq:deltaTop} cancel.

\subsection{Scalar potential, Einstein-Hilbert and scalar kinetic terms}
Since this calculation is similar to that for the usual EFTs, see for example the original discussion \cite{Hohm:2013vpa}, we will keep this section brief and mainly state results. However, we wish to emphasise again that because we will require integration by parts, the invariance under external diffeomorphisms only holds when the background trombone vanishes, i.e. $\tau_{ab} = 0$.

For the scalar potential, we calculate the anomalous variation under external diffeomorphisms of the first-order potential, \eqref{eq:SPot1Order}. Its anomalous variation is given by
\begin{equation}
 \begin{split}
  \Delta V &= \nabla_{cd} \xi^\mu \left[ \nabla_{ab} \gM^{ac} \D_\mu \gM^{bd} + \frac14 \gM^{ac} \gM^{bd} \nabla_{ab} \gM^{ef} \D_\mu \gM_{ef} - \gM^{ac} \gM^{be} \nabla_{ab} \gM^{df} \D_\mu \gM_{ef} \right. \\
  & \quad \left. + \frac12 \gM^{ac} \nabla_{ab} \gM^{bd} \D_\mu \ln g + \frac12 \gM^{ac} \nabla_{ab} \ln g \D_\mu \gM^{bd} + \frac14 \gM^{ac} \gM^{bd} \nabla_{ab} g^{\nu\rho} \D_\mu g_{\nu\rho} \right. \\
  & \quad \left. - \frac14 \gM^{ac} \gM^{bd} \nabla_{ab} \ln g \D_\mu \ln g \right] + \D_\mu \nabla_{cd} \xi^\mu \left( \gM^{ac} \nabla_{ab} \gM^{bd} - \frac12 \gM^{ac} \gM^{bd} \nabla_{ab} \ln g \right) \\
  & \quad + \D_\mu \nabla_{cd} \xi^\rho \left( \frac12 \gM^{ac} \gM^{bd} g_{\rho\nu} \nabla_{ab} g^{\mu\nu} \right) \,.
 \end{split}
\end{equation}
Let integrate this result by parts so that we only have $\nabla_{cd} \xi^\mu$ terms. We find
\begin{equation}
 \begin{split}
  \Delta V &= \nabla_{cd} \xi^\mu \left[ \frac14 \gM^{ac} \gM^{bd} \nabla_{ab} \gM^{ef} \D_\mu \gM_{ef} - \gM^{ac} \D_\mu \nabla_{ab} \gM^{bd} - \gM^{ac} \gM^{be} \nabla_{ab} \gM^{df} \D_\mu \gM_{ef} \right. \\
  & \quad \left. - \frac12 \gM^{ac} \nabla_{ab} \ln g \D_\mu \gM^{bd} + \frac14 \gM^{ac} \gM^{bd} \nabla_{ab} g^{\nu\rho} \D_\mu g_{\nu\rho} - \frac12 \gM^{ac} \gM^{bd} \D_\mu \nabla_{ab} \ln g \right] \\
  & \quad + \nabla_{cd} \xi^\rho \left[ - \frac14 \gM^{ac} \gM^{bd} g_{\rho\nu} \nabla_{ab} g^{\mu\nu} \D_\mu \ln g - \gM^{ac} g_{\rho\nu} \D_\mu \gM^{bd} \nabla_{ab} g^{\mu\nu} \right. \\
  & \quad \left. + \frac12 \gM^{ac} \gM^{bd} \D_\mu g^{\mu\nu} \nabla_{ab} g_{\rho\nu} + \frac12 \gM^{ac} \gM^{bd} \nabla_{ab} g_{\rho\nu} \D_\mu g^{\mu\nu} \right] \,.
 \end{split}
\end{equation}

We also need the variation of the Einstein-Hilbert term \eqref{eq:SEH}. It is given by
\begin{equation}
 \begin{split}
  \Delta_\xi R &= \frac12 \gM^{ac} \gM^{bd} \nabla_{ab} \xi^\lambda \left[ \nabla_{cd} \D_\lambda \ln g - g^{\mu\nu} \nabla_{cd} \D_\nu g_{\mu\lambda} - \frac12 \nabla_{cd} g_{\nu\rho} \D_\lambda g^{\nu\rho} \right. \\
  & \quad \left. - \frac12 g^{\mu\nu} \D_\nu \ln g \nabla_{cd} g_{\mu\lambda} - \nabla_{cd} g_{\mu\lambda} \D_\nu g^{\mu\nu} \right] - \frac12 g^{\mu\nu} g^{\rho\sigma} \D_\mu g_{\sigma\lambda} \Fa_{\rho\nu}{}^{ab} \nabla_{ab} \xi^\lambda \\
  & \quad - \frac12 g^{\mu\nu} \Fa_{\rho\mu}{}^{ab} \nabla_{ab} \D_\nu \xi^\rho \,.
 \end{split}
\end{equation}

Finally, we will need the variation of the scalar kinetic term. We integrate it by parts so that
\begin{equation}
 \begin{split}
 \delta_\xi L_{SK} &= g^{\mu\nu} \D_\mu \gM^{ab} \gM_{ac} \Fa_{\nu\rho}{}^{cd} \nabla_{bd} \xi^\rho + \nabla_{cd} \xi^\mu \left[ - \frac14 \D_\mu \gM^{ab} \nabla_{ef} \gM_{ab} \gM^{ce} \gM^{df} \right. \\
  & \left. \quad + \frac12 \D_\mu \gM^{ac} \gM^{bd} \nabla_{ab} \ln g + \D_\mu \gM^{ab} \nabla_{bf} \gM_{ae} \gM^{ce} \gM^{df} + \nabla_{ab} \D_\mu \gM^{ac} \gM^{bd} \right] \\
  & + \nabla_{cd} \xi^\rho \D_\mu \gM^{ac} \gM^{bd} g_{\nu\rho} \nabla_{ab} g^{\mu\nu}\,.
 \end{split}
\end{equation}
It is now a simple calculation to check that the anomalous variations of the different terms cancel.

\bibliographystyle{JHEP}
\bibliography{NewBib}

\end{document}